\documentclass[aps]{revtex4}
\usepackage{amsfonts}
\usepackage{amsmath}
\usepackage{amssymb,epsf}

\begin{document}

\title{Neutron stars structure in the context of massive gravity}
\author{S. H. Hendi$^{1,2}$\footnote{%
email address: hendi@shirazu.ac.ir}, G. H. Bordbar$^{1,3}$\footnote{%
email address: ghbordbar@shirazu.ac.ir}, B. Eslam Panah$^{1,4}$\footnote{%
email address: behzad.eslampanah@gmail.com} and S. Panahiyan$^{1,5}$ \footnote{%
email address: sh.panahiyan@gmail.com} } \affiliation{$^1$ Physics
Department and Biruni Observatory, College of Sciences, Shiraz
University, Shiraz 71454, Iran\\
$^2$ Research Institute for Astronomy and Astrophysics of Maragha (RIAAM),
P. O. Box 55134-441, Maragha, Iran\\
$^3$ Center for Excellence in Astronomy and Astrophysics
(CEAA-RIAAM)-Maragha, P. O. Box 55134-441, Maragha, Iran\\
$^4$ ICRANet, Piazza della Repubblica 10, I-65122 Pescara, Italy\\
$^5$ Physics Department, Shahid Beheshti University, Tehran 19839,
Iran}

\begin{abstract}
Motivated by the recent interests in spin$-2$ massive gravitons,
we study the structure of neutron star in the context of massive
gravity. The modifications of TOV equation in the presence of
massive gravity are explored in $4$ and higher dimensions. Next,
by considering the modern equation of state for the neutron star
matter (which is extracted by the lowest order constrained
variational (LOCV) method with the AV$18$ potential), different
physical properties of the neutron star (such as Le Chatelier's
principle, stability and energy conditions) are investigated. It
is shown that consideration of the massive gravity has specific
contributions into the structure of neutron star and introduces
new prescriptions for the massive astrophysical objects. The
mass-radius relation is examined and the effects of massive
gravity on the Schwarzschild radius, average density, compactness,
gravitational redshift and dynamical stability are studied.
Finally, a relation between mass and radius of neutron star versus
the Planck mass is extracted.
\end{abstract}

\maketitle

\section{Introduction}

The Einstein theory of gravity has been a pioneering tool for
understanding and describing the gravitational systems. Most of
the results and validations of this theory have been confirmed by
observations done on solar system level. In addition, the results
of LIGO proved the existence of gravitational wave which was one
of the challenging predictions of general relativity (GR)
\cite{LIGO}. It is expected to observe the advantages of GR for
beyond the Newtonian theory regimes (high curvature regimes) such
as near the compact objects. The existence of compact objects has
been confirmed in the Einstein theory. Due to the physical
properties of these objects, it is necessary to take the curvature
of spacetime into account in order to have reliable predictions.
Therefore, we employ the Einstein theory of gravity to study the
neutron star.

Einstein theory predicts the existence of massless spin-%
$2$ gravitons with two degrees of freedom as intermediate particles for
signaling the gravitational interactions. But, there have been several
arguments regarding the possibility of existence of massive gravitons. These
arguments are supported by studies that are conducted on quantum level of
gravity and brane-world gravity. Especially problems such as the hierarchy
problem and their brane-world gravity solutions have expressed on the
possibility of existence of massive spin-$2$ gravitons \cite%
{DvaliGP,DvaliGPI}. Therefore, it is natural to relax massless constraint
and consider the modifications and generalizations of the general relativity
to include massive graviton. In this paper, we generalize the Einstein
theory of gravity to include the massive graviton and investigate its
effects on the hydrostatic equilibrium equation of a typical neutron star.

The first attempt for constructing the massive gravity was done by
Fierz and Pauli \cite{FierzI}. This theory has the specific
problem known as vDVZ (van Dam-Veltman-Zakharov) discontinuity
which indicates that the propagators of
massless and massive in the limit of $m\rightarrow 0$, are not the same \cite%
{van Dam,Zakharov,DeserW}. One of the resolutions of this problem
was Vainshtein mechanism which requires the system to be
considered in the nonlinear framework \cite{Vainshtein} (it is
notable that in nonlinear dRGT, there are also vacua that are free
from vDVZ discontinuity \cite{Zhou1}). Such generalization to
nonlinear case introduces a ghost into the theory which is known
as Boulware-Deser ghost \cite{Boulware}. There are various ghost
free scenarios for considering the massive gravity in the
nonlinear framework. One of the interesting ghost free theories of
massive gravity is known as dRGT theory which was developed by de
Rham, Gabadadze and Tolley \cite{de RhamGT,Hinterbichler}. In this
theory, a reference metric is employed to build massive terms
\cite{de RhamGT,Hinterbichler,HassanRI,HassanRS}. These massive
terms are inserted in the action to provide massive gravitons.
Cosmological results, black hole solutions and their
thermodynamical properties in this massive gravity are
investigated by many authors
\cite{Fasiello,CaiEGS,BambaHMNS,CaiS,GoonGH,HeisenbergKY,SolomonEA,PanC,KodamaA,BabichevF,BabichevB,SureshMP,LiLX,Zhou2,Zhou3,Zou}.
In addition, Katsuragawa et al, studied the neutron stars in the
context of dRGT theory and showed that, the massive gravity leads
to small deviation from the GR \cite{Katsuragawa}.

It is worthwhile to mention that the reference metric plays a key role for
constructing the massive theory of gravity \cite{review}. One of the
modifications in reference metric was done by Vegh which introduced a new
massive gravity \cite{Vegh}. This new massive theory has specific
applications in the gauge/gravity duality especially in lattice physics
which motivate one to use it in other frameworks as well. This theory was
employed in the context of black holes and it was shown that geometrical and
thermodynamical structures of the black holes will be modified and new
phenomena were reported for massive black holes \cite%
{CaiHPZ,HendiEPJHEP2015,XuCH,GhoshTW,HendiPEJHEP,HendiEPJHEP2016,SureshK,Upadhyay,Rappid}.
Here, we use this massive theory to conduct our studies in the
properties of neutron star.

The structure of stars and their phenomenological properties are
described with hydrostatic equilibrium equation (HEE). This
equation is based on the fact that a typical star will be in
equilibrium when there is a balance between the internal pressure
and the gravitational force. Historically speaking, the first HEE
equation for GR was introduced and employed by Tolman, Oppenheimer
and Volkoff (TOV) \cite{Tolman,Tolman1939,Oppenheimer}. After
that, a series of studies were dedicated to obtain HEE of neutron
star
\cite{YunesV,SilbarR,NarainSM,BordbarH,BoonsermVW,LiWC,OliveiraVFS,HeFLN}.
In addition, the compact objects and their TOV equations were
investigated in the presence of different modified gravities such
as; gravity's rainbow \cite{HendiBEP}, vector-tensor-Horndeski
theory of gravity \cite{MomeniarXiv}, dilaton gravity
\cite{Hendi2015}, $F(R)$ and $F(G)$ gravities
\cite{CapozzielloF(R),AstashenokCOU,AstashenokCOJCAP,SavasDY,ZhouYZY}
(see
\cite{Harada,Wiseman,Sotani,DelidumanEK,DonevaYSK,ChamelHZF,BaraussePPL,SilvaSBH,AsCoOd,BrihayeR,DasRGR,
StaykovDYK,Boyadjiev,Meyer,OrellanaGPR,ArbanilLZ,GoswamiNMG,GlampedakisPSB,GreenS,CandelasHSW}
for more details).

According to recent studies on the neutron stars and observations
of the interesting properties of them
\cite{Sigurdsson,Vietri,Heger,Rea,Perna,Katayama,Clausen,Hebeler,
Perez,Deaton,Dall'Osso,Venumadhav,Poutanen,East,Zheng,Mastrano,Giacomazz,Gelfand,Wielgus,Ruiz},
we want to investigate these stars in the context of massive
gravity. In other words, our main motivation in this paper is
studying the effects of considering the massive gravity on the
structure of neutron stars. Previous studies in the context of
other astrophysical objects have proven a wide variation in the
properties of these objects comparing to the massless graviton
case. Therefore, we are expecting to see the specific
modifications in the properties of neutron star as well. Here, we
would like to address how the structure of neutron star will be
modified in the presence of massive gravity and which
contributions this generalization has into properties of these
objects. We regard the HEE equation in $4$ and higher dimensions
with a suitable equation of state (EoS), which satisfies
stability, energy conditions and Le Chatelier's principle. We
obtain the maximum mass and corresponding radius, Schwarzschild
radius, compactness, gravitational redshift and dynamical
stability of the neutron stars. We give details regarding the
effects of massive gravity on these properties. These studies
provide an insight into the structure of neutron stars and enable
one to make a comparison between the massive and massless gravity
theories. Remembering that the neutron stars, similar to other
massive objects, propagate the gravitational waves, one is urged
to study the neutron stars in the presence of massive gravity
which is the aim of this paper.

The outline of our paper is as follows. In Sec. \ref{ModTOV}, we consider a
spherical symmetric metric and obtain the modified TOV in Einstein-massive
gravity in four dimensions. Next, we employ the specific many-body EoS and
study its properties such as the Le Chatelier's principle, stability and
energy conditions. Then, considering the Einstein-massive gravity, we
investigate the neutron star structure and obtain other properties of this
star. In next section, we extract mass and radius of this star versus the
Planck mass as a fundamental physical constant. Finally, we finish our paper
with some closing remarks.

\section{Modified TOV equation in the massive gravity}\label{ModTOV}

The action of Einstein-massive (EN-massive) gravity with the cosmological
constant in $d$-dimensions is given by
\begin{equation}
\mathcal{I}=-\frac{1}{16\pi }\int d^{d}x\sqrt{-g}\left[ \mathcal{R}-2\Lambda
+m^{2}\sum_{i}^{4}c_{i}\mathcal{U}_{i}(g,f)\right] +I_{matter},
\label{Action}
\end{equation}%
where $R$ and $m$ are the Ricci scalar and the massive parameter, $\Lambda $
is the negative cosmological constant, and $f$ and $g$ are a fixed symmetric
tensor and metric tensor, respectively. In addition, $c_{i}$'s are constants
and $\mathcal{U}_{i}$'s are symmetric polynomials of the eigenvalues of $%
d\times d$ matrix $K_{\nu }^{\mu }=\sqrt{g^{\mu \alpha }f_{\alpha \nu }}$
where they can be written in the following forms
\begin{eqnarray}
\mathcal{U}_{1} &=&\left[ \mathcal{K}\right] ,\;\;\;\;\;\;\;\mathcal{U}_{2}=%
\left[ \mathcal{K}\right] ^{2}-\left[ \mathcal{K}^{2}\right] ,  \nonumber \\
\mathcal{U}_{3} &=&\left[ \mathcal{K}\right] ^{3}-3\left[ \mathcal{K}\right] %
\left[ \mathcal{K}^{2}\right] +2\left[ \mathcal{K}^{3}\right] ,  \nonumber \\
\mathcal{U}_{4} &=&\left[ \mathcal{K}\right] ^{4}-6\left[ \mathcal{K}^{2}%
\right] \left[ \mathcal{K}\right] ^{2}+8\left[ \mathcal{K}^{3}\right] \left[
\mathcal{K}\right] +3\left[ \mathcal{K}^{2}\right] ^{2}-6\left[ \mathcal{K}%
^{4}\right] .  \nonumber
\end{eqnarray}

By variation of Eq. (\ref{Action}) with respect to the metric tensor $g_{\mu
}^{\nu }$, the equation of motion for EN-massive gravity can be written as%
\begin{equation}
G_{\mu }^{\nu }+\Lambda g_{\mu }^{\upsilon }+m^{2}\chi _{\mu }^{\upsilon
}=K_{d}T_{\mu }^{\nu },  \label{Field equation}
\end{equation}%
where $K_{d}=\frac{8\pi G_{d}}{c^{4}}$, $G_{d}$ is $d$-dimensional
gravitational constant, $G_{\mu \nu }$ is the Einstein tensor and $c$ is the
speed of light in vacuum. Also, $T_{\mu }^{\nu }$ denotes the
energy-momentum tensor which comes from the variation of $I_{matter}$ and $%
\chi _{\mu \nu }$ is the massive term with the following explicit form
\begin{eqnarray}
\chi _{\mu \nu } &=&-\frac{c_{1}}{2}\left( \mathcal{U}_{1}g_{\mu \nu }-%
\mathcal{K}_{\mu \nu }\right) -\frac{c_{2}}{2}\left( \mathcal{U}_{2}g_{\mu
\nu }-2\mathcal{U}_{1}\mathcal{K}_{\mu \nu }+2\mathcal{K}_{\mu \nu
}^{2}\right)  \nonumber \\
&&  \nonumber \\
&&-\frac{c_{3}}{2}(\mathcal{U}_{3}g_{\mu \nu }-3\mathcal{U}_{2}\mathcal{K}%
_{\mu \nu }+6\mathcal{U}_{1}\mathcal{K}_{\mu \nu }^{2}-6\mathcal{K}_{\mu \nu
}^{3})  \nonumber \\
&&  \nonumber \\
&&-\frac{c_{4}}{2}\left( \mathcal{U}_{4}g_{\mu \nu }-4\mathcal{U}_{3}%
\mathcal{K}_{\mu \nu }+12\mathcal{U}_{2}\mathcal{K}_{\mu \nu }^{2}-24%
\mathcal{U}_{1}\mathcal{K}_{\mu \nu }^{3}+24\mathcal{K}_{\mu \nu
}^{4}\right) .
\end{eqnarray}

\subsection{Modified TOV equation in (3+1)-dimensions}

In this section, the static solutions of EN-massive gravity in $(3+1)$%
-dimensions are obtained. For this purpose, we consider a
spherical symmetric space-time in the following form
\begin{equation}
g_{\mu \nu }=diag\left( f(r),-g(r)^{-1},-r^{2},-r^{2}\sin ^{2} \theta
\right) ,  \label{Metric}
\end{equation}%
where $f(r)$ and $g(r)$ are unknown metric functions. Now, in
order to obtain exact solutions, we should consider a suitable
reference metric. Obeying the ansatz of Ref. \cite{CaiHPZ}, we
consider the following relation for the reference metric
\begin{equation}
f_{\mu \nu }=diag(0,0,C^{2}r^{2},C^{2}r^{2}\sin ^{2} \theta ),  \label{f11}
\end{equation}
in which $C$ is a positive constant. Considering the metric ansatz
(\ref{f11}), we can obtain the explicit forms of nonzero
$\mathcal{U}_{i}$'s as \cite{CaiHPZ}
\begin{equation}
\mathcal{U}_{1}=\frac{2C}{r},\text{ \ \
}\mathcal{U}_{2}=\frac{2C^{2}}{r^{2}}.  \nonumber
\end{equation}

Here, we regard the neutron star as a perfect fluid with the following
energy-momentum tensor
\begin{equation}
T^{\mu \nu }=\left( c^{2}\rho +P\right) U^{\mu }U^{\nu }-Pg^{\mu \nu },
\label{EMTensorEN}
\end{equation}
where $P$ and $\rho $ are the pressure and density of the fluid which are
measured by the local observer, respectively, and $U^{\mu }$ is the fluid
four-velocity. Using Eqs. (\ref{Field equation}) and (\ref{EMTensorEN}) with
the metric introduced in Eq. (\ref{Metric}), it is easy to obtain the
components of energy-momentum as
\begin{equation}
T_{0}^{0}=\rho c^{2}~\ \ \ \ \&~\ \ \ \ T_{1}^{1}=T_{2}^{2}=T_{3}^{3}=-P.
\label{4dim}
\end{equation}

In addition, taking into account Eqs. (\ref{Metric}) and (\ref{4dim}), it is
straightforward to achieve the following nonzero components of field
equation (\ref{Field equation})
\begin{eqnarray}
Kc^{2}r^{2}\rho &=&\Lambda r^{2}+\left( 1-g\right) -r g{^{\prime }-m}%
^{2}C\left( c_{1}r+c_{2}C\right) ,  \label{1} \\
&&  \nonumber \\
Kfr^{2}P &=&-\Lambda r^{2}f-\left( 1-g\right) f+rg{f{^{\prime }}+m}%
^{2}fC\left( c_{1}r+c_{2}C\right) {,}  \label{2} \\
&&  \nonumber \\
4Kf^{2}r P &=&-4\Lambda rf^{2}+2\left( gf\right) {^{\prime }}f-r g f{^{\prime 2}%
}+r\left[ g{^{\prime }}f{^{\prime }+2g}f{^{\prime \prime }}\right]
f+2m^{2}Cc_{1}f^{2},  \label{33}
\end{eqnarray}%
where $K=\frac{8\pi G}{c^{4}}$, and $f$, $g$, $\rho $ and $P$ are functions
of $r$. Also, we note that the prime and double prime denote the first and
the second derivatives with respect to $r$, respectively.

Using Eqs. (\ref{1})-(\ref{33}) and after some calculations, we obtain
\begin{equation}
\frac{dP}{dr}+\frac{\left( c^{2}\rho +P\right) f{^{\prime }}}{2f}=0.
\label{extraEQ}
\end{equation}

In addition, for extracting $g(r)$, we can use Eq. (\ref{1}) which leads to
\begin{equation}
g(r)=1+\frac{\Lambda }{3}r^{2}-m^{2}C\left( \frac{c_{1}r}{2}+c_{2}C\right) -%
\frac{c^{2}KM(r)}{4\pi r},  \label{4g(r)}
\end{equation}%
in which $M(r)=\int 4\pi r^{2}\rho (r)dr$. Now, we obtain
$f^{\prime }$ from Eq. (\ref{2}) and insert it with Eq.
(\ref{4g(r)}) in Eq. (\ref{extraEQ}) to calculate HEE in the
Einstein-massive gravity with the following form
\begin{equation}
\frac{dP}{dr}=\frac{\left( c^{2}\rho +P\right) \left[ \frac{c^{2}KM(r)}{2}%
+2\pi r^{3}\left( \frac{2\Lambda }{3}+KP\right) -m^{2}\pi r^{2}c_{1}C\right]
}{r\left[ 2m^{2}\pi r^{2}c_{1}C+c^{2}KM(r)+4\pi r\left( m^{2}c_{2}C^{2}-%
\frac{\Lambda }{3}r^{2}-1\right) \right] },  \label{TOV}
\end{equation}%
which is modified TOV equation due to the presence of massive graviton. As
one expects, for $m=0$, Eq. (\ref{TOV}) is reduced to the following TOV
equation obtained in Einstein-$\Lambda $ gravity \cite{TOVLambda,TOV-Lambda}
\begin{equation}
\frac{dP}{dr}=\frac{\left[ 3c^{2}G M(r)+r^{3}\left( \Lambda
c^{4}+12\pi
GP\right) \right] }{c^{2}r\left[ 6 G M(r)-c^{2}r\left( \Lambda r^{2}+3\right) %
\right] }\left( c^{2}\rho +P\right) .  \label{4TOV}
\end{equation}

In addition, in the absence of both massive term and cosmological constant ($%
m=\Lambda =0$), Eq. (\ref{TOV}) leads to the usual TOV equation of
Einstein gravity (see \cite{Tolman,Tolman1939,Oppenheimer} for
more details). It is notable that, the generalization to higher
dimensions is done in the appendix A.

Before applying the mentioned gravitational framework on the
neutron star structure, we should point out some comments for the
mentioned massive gravity. As we mentioned before, the massive
gravity employed in this paper is essentially a dRGT like \cite
{review}. It was shown that all dRGT like theories of the massive
gravity in $d$ dimensions with $N$ scalar fields (St\"{u}ckelberg
fields) enjoy at most $\frac{1}{2}d\left( d-3\right) +N$ number of
degrees of freedom. The reference metric employed in this paper is
spatial reference metric which in its simpler form in the
appropriate orthonormal coordinate, it will be $( 0,0,1,1)$. This
specific choice in reference metric leads into interesting results
which among them one can point out that under certain coordinate
transformation, the general covariance is preserved in radial and
temporal coordinates while it breaks in spatial dimensions
\cite{Vegh}. Therefore, this theory indeed enjoys Lorentz
violating property. On the other hand, it was shown that this
choice of reference metric and more general ones enjoy the
preservation of the Hamiltonian constraint which leads to removing
one of the degrees of freedom. In addition, based on the
diffeomorphism, another degree of freedom is eliminated. In
general, since this is a $4$ dimensional theory of massive
gravity, there might be up to $6$ degrees of freedom. Two of these
degrees of freedom were eliminated due to mentioned properties. On
the other hand, it was also shown that only two St\"{u}ckelberg
fields exist in the diffeomorphism invariant formulation of this
theory \cite{Vegh}. This indicates that two degrees of freedom are
absent which leads to absence of Boulware-Deser (BD) ghost.

To summarize, it is notable that depending on the notion of time,
the number of degrees of freedom can differ from two to five.
Although an observer with arbitrary time has to describe five
degrees of freedom, the observers with St\"{u}ckelberg time
function will describe two degrees of freedom such as the
situation of massless gravitons in the usual general relativity.

The full details regarding absence of the BD ghost in this massive
gravity are given in Refs \cite{Vegh,Chin}. Especially in Ref.
\cite{Chin}, the stability of massive gravity with singular metric
of arbitrary rank was studied and the absence of BD ghost was
proven.

Evaluating the second derivatives of the massive action with
respect to the background graviton field, one can obtain a mass
matrix. In this case, the eigenvalues of the matrix would
correspond to the masses of each mode, the so-called, $m_{tt}$,
$m_{ij}$, etc. These are the mass fluctuations of the modes which
are depending on the non-trivial contributions of the
St\"{u}ckelberg function. In other words, such mass fluctuations
are depending on the free parameters of the massive theory such as
$c_{1}$ and $c_{2}$, etc. So, it would be useful to calculate the
physical mass of fluctuations and impose appropriate conditions to
avoid tachyon-like instabilities. Taking into account the point of
Ref. \cite{review}, we can regard that the mass parameter is of
the order of the Hubble parameter today, and therefore, such an
instability would not be problematic. However, since the
tachyon-like instabilities are very important in some
gravitational framework, such as black holes, we will address such
substantial point in an independent paper.

\section{Structure properties of neutron star \label{Structure}}

\subsection{Equation of state of neutron star matter}

The interior region of a typical neutron star is a mixed soup of neutrons,
protons, electrons and muons in charge neutrality and beta equilibrium
conditions (beta-stable matter) \cite{Shapiro}. This balanced mixture is
governed by unknown EoS. One of the EoS which could be employed to study the
neutron star is the microscopic constrained variational calculations based
on the cluster expansion. This EoS has been employed to study the structure
of neutron star matter before \cite{Bordbar,Hendi2015}. Fundamentally, the
mentioned model is based on two-nucleon potentials which are the modern
Argonne AV$18$ \cite{Wiringa} and charged dependent Reid-$93$ \cite{Stoks}.
It is notable that this method requires no free parameter, has a good
convergence and is more accurate comparing to other semi-empirical parabolic
approximation methods. These advantages come from a microscopic computation
of asymmetry energy which is carried on for the asymmetric nuclear matter
calculations. The necessity of microscopic calculations with the modern
nucleon-nucleon potentials which is isospin projection ($T_{z}$) dependent
was pointed out in Ref. \cite{Bordbar1998}. Here, we employ the lowest order
constrained variational (LOCV) method with the AV$18 $ potential \cite%
{Bordbar} for obtaining the modern EoS for neutron star matter and
investigating some physical properties of neutron star structure.

As we mentioned, the energy of the system under study is obtained by the
LOCV method which is a fully self-consistent formalism. Through a
normalization constraint, this method keeps the higher order terms as small
as possible \cite{Owen}. In addition, this method has been employed to
calculate the properties of neutron, nuclear and asymmetric nuclear matters
at zero and finite temperatures \cite{Owen,Modarres,Howes}. The functional
minimization procedure represents an enormous computational simplification
over the unconstrained methods which attempt to go beyond the lowest order.

\begin{figure}[tbp]
$%
\begin{array}{c}
\epsfxsize=7cm \epsffile{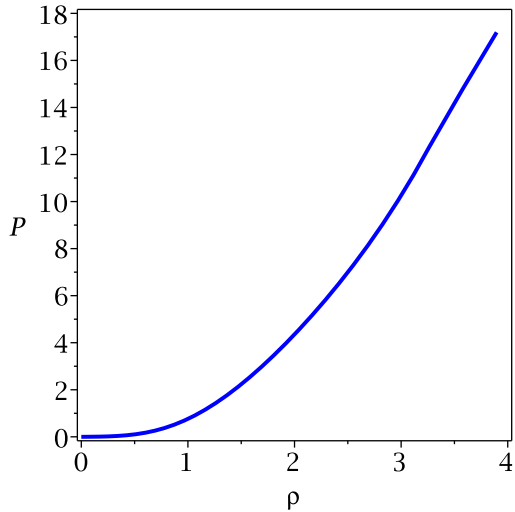}%
\end{array}
$%
\caption{Equation of state of neutron star matter (pressure, $P$ ($10^{35}$
erg/$cm^{3}$) versus density, $\protect\rho $ ($10^{15}$ g/$cm^{3}$)).}
\label{Fig1}
\end{figure}

A trial many-body wave function is
\begin{equation}
\psi =F\phi ,
\end{equation}%
where $\phi $ is the uncorrelated ground-state wave function of $N$
independent neutrons, and $F$ is a proper $N$-body correlation function.
Here, we apply Jastrow approximation \cite{Jastrow} to replace $F$ as
\begin{equation}
F=S\prod_{i>j}f(ij),
\end{equation}%
where $S$ and $f(ij)$ are a symmetrizing operator and the two-body
correlation function, respectively. Besides, we consider a cluster expansion
of the energy functional up to the two-body term
\begin{equation}
E([f])=\frac{1}{N}\frac{\langle \psi |H|\psi \rangle }{\langle \psi |\psi
\rangle }=E_{1}+E_{2},
\end{equation}%
in which $\psi $ and $H$ are wave function and Hamiltonian system,
respectively. In other words, the energy per particle up to the two-body
term is
\begin{equation}
E([f])=E_{1}+E_{2},
\end{equation}%
where $E_{1}=\sum_{i=+,-}\frac{3}{5}\frac{\hbar ^{2}k_{F}^{(i)^{2}}}{2m}%
\frac{\rho ^{(i)}}{\rho }$ and $E_{2}=\frac{1}{2N}\sum_{ij}\langle ij|\nu
(12)|ij-ji\rangle $ are one-body and two-body energy terms, respectively. It
is notable that, $k_{F}^{(i)}=\left( 6\pi ^{2}\rho ^{(i)}\right) ^{1/3}$is
the Fermi momentum of a neutron with spin projection $i$. The operator $\nu
(12)$ is nuclear potential and it has been given in Ref. \cite{Bordbar044310}
(see Refs. \cite{Bordbar1998} for more details). The behavior of obtained
EoS of neutron star matter is shown in Fig. \ref{Fig1}. We extract the
mathematical forms for the EoS presented in Fig. \ref{Fig1} as
\begin{equation}
P=\sum_{i=1}^{7}\mathcal{A}_{i}\rho ^{7-i},  \label{EoS}
\end{equation}%
in which $\mathcal{A}_{i}$ are%
\begin{eqnarray*}
\mathcal{A}_{1} &=&-3.518\times 10^{-57},\ \ \ \ \ \mathcal{A}%
_{2}=3.946\times 10^{-41},\ \ \ \ \mathcal{A}_{3}=-1.67\times 10^{-25},\  \\
\ \mathcal{A}_{4} &=&3.242\times 10^{-10},\ \ \ \ \ \ \ \mathcal{A}%
_{5}=-1.458\times 10^{5},\ \ \ \ \ \mathcal{A}_{6}=2.911\times 10^{19}, \\
\mathcal{A}_{7} &=&-9.983\times 10^{31}.
\end{eqnarray*}

In order to investigate the properties of such EoS with more details, we
study Le Chatelier's principle condition in the following subsection.

\subsubsection{Le Chatelier's principle}

The matter of {star satisfies $dP/d\rho \geq 0$ which is a necessary
condition of a stable body both as a whole and also with respect to the
non-equilibrium elementary regions with spontaneous contraction or expansion
(Le Chatelier's principle) \cite{Glendenning}. As one can see, in Fig. \ref%
{Fig1}, Le Chatelier's principle is established. }

The stability ($0\leq v^{2}=\left( \frac{dP}{d\rho }\right) \leq c^{2}$) and
energy conditions for this EoS are investigated in Ref. \cite{HendiBEP}, and
it is shown that this EoS satisfied these conditions.

\subsection{Mass-radius relation and other properties of neutron star in
massive gravity}

Considering the maximum gravitational mass of a neutron star for dynamical
stability against gravitational collapse into a black hole, one is able to
make differences between neutron star and black holes. In other words, there
is a critical maximum mass for the massive object in which for masses larger
than the maximum value, the massive object becomes a black hole \cite%
{Shapiro}. The value of maximum mass originated from the nucleons degeneracy
pressure is evidently the possible maximum mass of neutron star. Therefore,
obtaining the maximum gravitational mass of neutron star is of a great
interest, and important in astrophysics. Unfortunately, the advanced
observational technologies for measuring the mass of neutron star by
investigating the X-ray pulsars and X-ray bursters were not able to produce
accurate results. Nevertheless the measurements that are done with the
binary radio pulsars \cite{WeisbergT,Liang,HeapC,Quaintrell}, provided
highly accurate results for the mass of neutron star. In Ref. \cite{BordbarH}%
, the Einstein gravity has been investigated, and maximum mass of neutron
star has been obtained using the modern equations of state of neutron star
matter obtained from the microscopic calculations. It was shown that the
maximum mass of neutron star is about $1.68M_{\odot }$. In addition, the EoS
with dilaton gravity was employed and the properties of neutron star were
investigated \cite{Hendi2015}. The results showed that by increasing the
effects of dilaton gravity, the maximum mass of this star decreases ($%
M_{\max }\leq 1.68M_{\odot }$). Here, we intend to obtain the maximum mass
of neutron star by considering the obtained TOV equation for
Einstein-massive gravity (Eq. (\ref{TOV})) and investigate the properties of
neutron star.

\begin{figure}[tbp]
$%
\begin{array}{cc}
\epsfxsize=7cm \epsffile{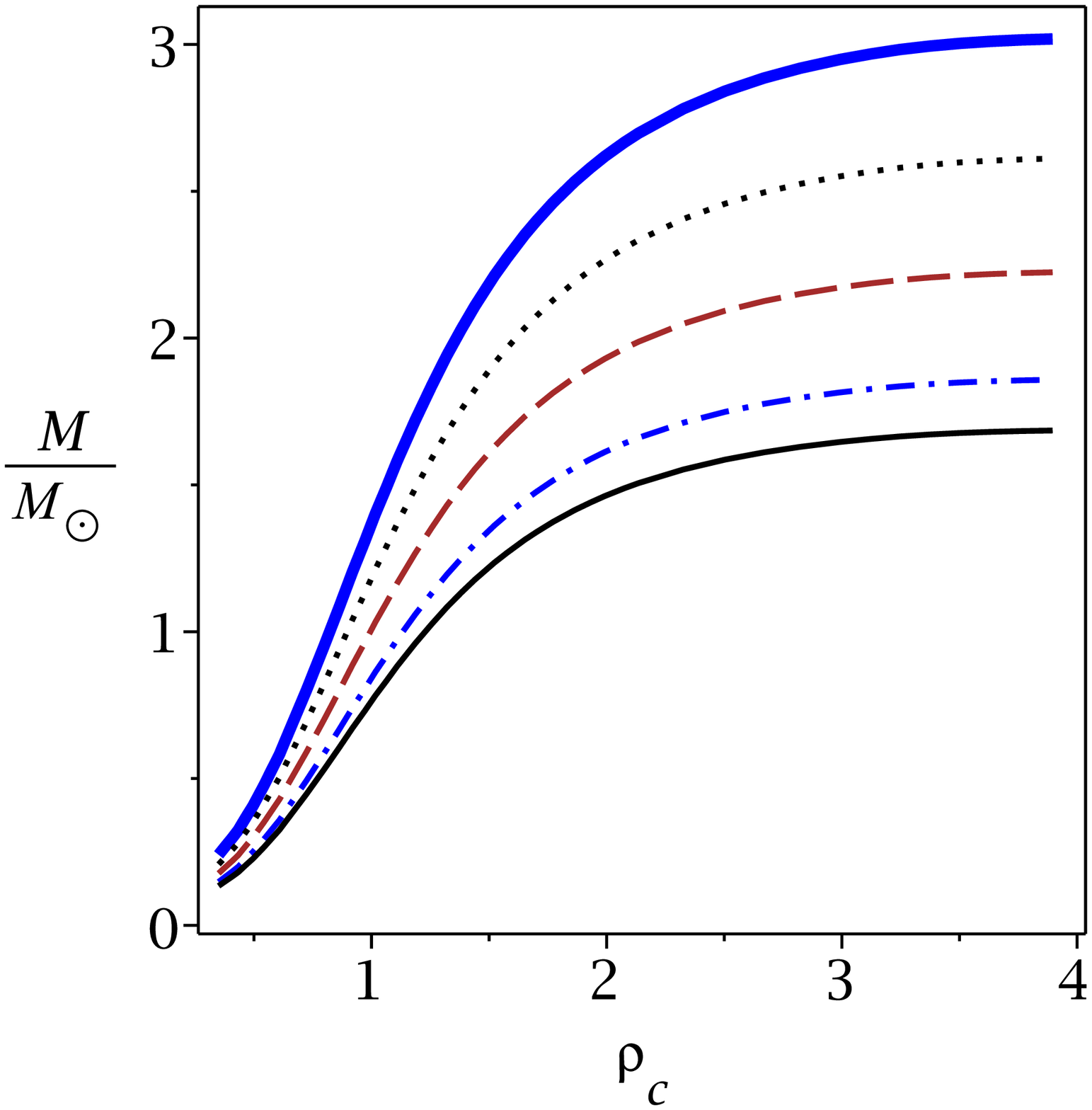} & \epsfxsize=7cm %
\epsffile{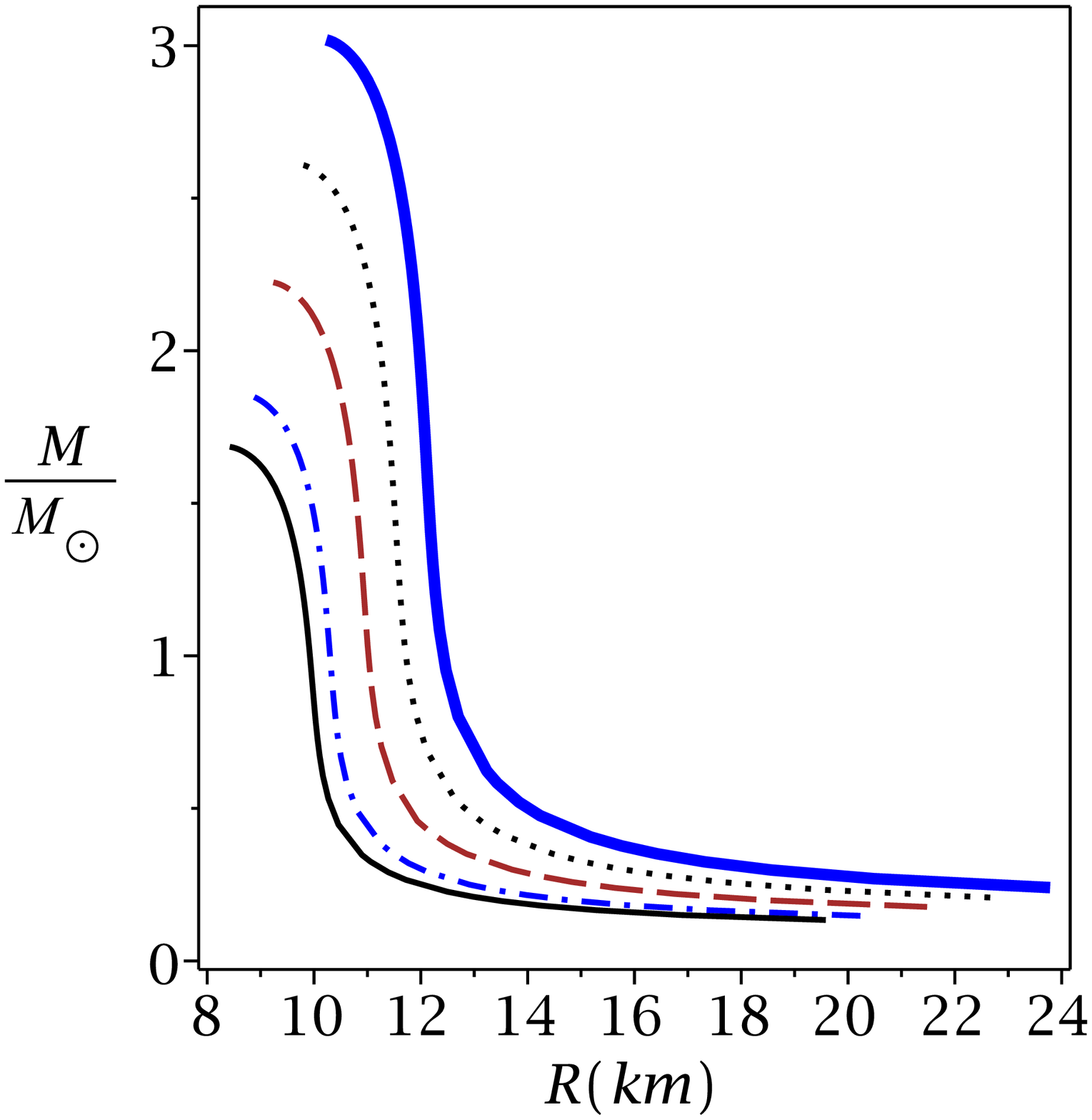}%
\end{array}
$%
\caption{Gravitational mass versus central mass density (radius), $\protect%
\rho _{c}$ ($10^{15}$gr/$cm^{3}$), for $C=2$ and $m^{2}c_{1}=3.168%
\times10^{-13}$.\newline
Left diagrams: gravitational mass versus central mass density for $%
m^{2}c_{2}=-1.204\times 10^{-1}$ (bold line), $m^{2}c_{2}=-8.238\times
10^{-2}$ (doted line), $m^{2}c_{2}=-5.386\times 10^{-2}$ (dashed line), $%
m^{2}c_{2}=-1.647\times 10^{-2}$ (dashed-dotted line) and $%
m^{2}c_{2}=-1.743\times 10^{-3}$ (continuous line).\newline
Right diagrams: gravitational mass versus radius for $m^{2}c_{2}=-1.204%
\times 10^{-1}$ (bold line), $m^{2}c_{2}=-8.238\times 10^{-2}$ (doted line),
$m^{2}c_{2}=-5.386\times 10^{-2}$ (dashed line), $m^{2}c_{2}=-1.647\times
10^{-2}$ (dashed-dotted line) and $m^{2}c_{2}=-1.743\times 10^{-3}$
(continuous line).}
\label{Fig3}
\end{figure}

\begin{figure}[tbp]
$%
\begin{array}{cc}
\epsfxsize=7cm \epsffile{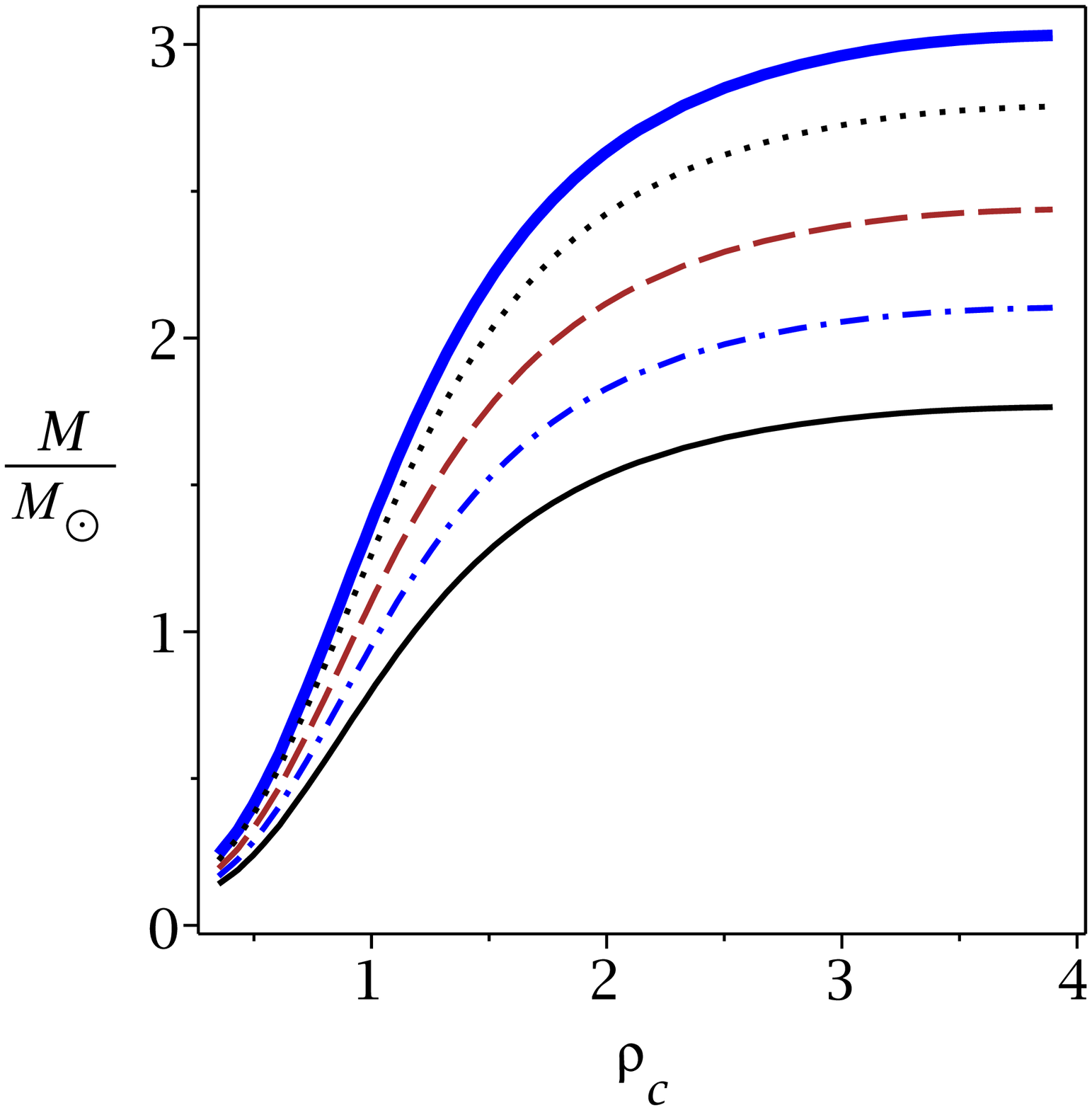} & \epsfxsize=7cm %
\epsffile{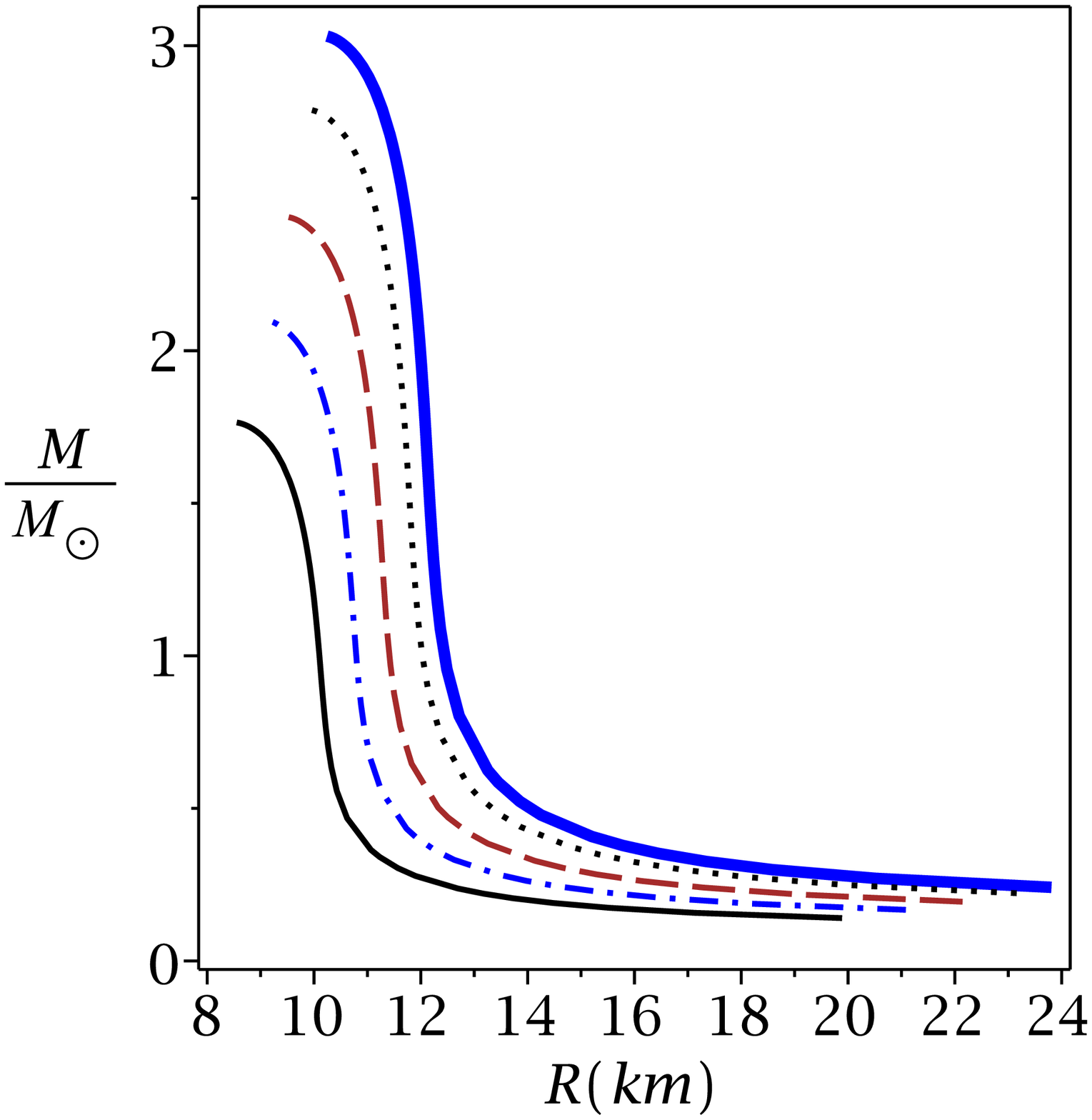}%
\end{array}
$%
\caption{Gravitational mass versus central mass density (radius), $\protect%
\rho _{c}$ ($10^{15}$gr/$cm^{3}$), for $m^{2}c_{1}=3.168\times 10^{-12}$ and
$m^{2}c_{2}=-3.168\times 10^{-2}$.\newline
Left diagrams: gravitational mass versus central mass density for $C=1.20$
(bold line), $C=2.32$ (doted line), $C=3.08$ (dashed line), $C=3.65$
(dashed-dotted line) and $C=3.85$ (continuous line). \newline
Right diagrams: gravitational mass versus radius for $C=1.20$ (bold line), $%
C=2.32$ (doted line), $C=3.08$ (dashed line), $C=3.65$ (dashed-dotted line)
and $C=3.85$ (continuous line).}
\label{Fig4}
\end{figure}

Now, by employing the EoS of neutron star matter presented in Fig. \ref{Fig1}
and numerical approach for integrating the HEE obtained in Eq. (\ref{TOV}),
we can calculate the maximum mass and other properties of the neutron star.
To do so, one can consider the boundary conditions $P(r=0)=P_{c}$ and $%
m(r=0)=0$, and integrates Eq. (\ref{TOV}) outwards to a radius
$r=R$ in which $P$ vanishes for selecting a $\rho_{c}$. This leads
to the neutron star radius $R$ and mass $M=m(R)$. We present the
results in different
figures and tables (see Figs. \ref{Fig3} and \ref{Fig4}, and tables \ref%
{tab1}, \ref{tab2} and \ref{tab3} for more details).
\begin{table*}[tbp]
\caption{Structure properties of neutron star in massive gravity for $C=2$
and $m^{2}c_{1}=3.168\times 10^{-5}$.}
\label{tab1}
\begin{center}
\begin{tabular}{ccccccc}
\hline\hline
$m^{2}c_2$ & ${M_{max}}\ (M_{\odot})$ & $R\ (km)$ & $R_{Sch}\ (km)$ & $%
\overline{\rho }$ $(10^{14}g$ $cm^{-3})$ & $\sigma (10^{-1})$ & $z(10^{-1})$
\\ \hline\hline
$-3.168\times 10^{-5}$ & $1.68$ & $8.42$ & $4.95$ & $13.36$ & $5.88$ & $5.58$
\\ \hline
$-3.168\times 10^{-4}$ & $1.68$ & $8.42$ & $4.95$ & $13.36$ & $5.88$ & $5.57$
\\ \hline
$-3.168\times 10^{-3}$ & $1.71$ & $8.47$ & $4.97$ & $13.36$ & $5.87$ & $5.57$
\\ \hline
$-1.584\times 10^{-2}$ & $1.84$ & $8.68$ & $5.10$ & $13.36$ & $5.87$ & $5.57$
\\ \hline
$-3.168\times 10^{-2}$ & $2.01$ & $8.94$ & $5.26$ & $13.36$ & $5.88$ & $5.58$
\\ \hline
$-6.337\times 10^{-2}$ & $2.36$ & $9.43$ & $5.56$ & $13.36$ & $5.89$ & $5.59$
\\ \hline
$-9.505\times 10^{-2}$ & $2.73$ & $9.89$ & $5.83$ & $13.40$ & $5.89$ & $5.61$
\\ \hline
$-1.267\times 10^{-1}$ & $3.11$ & $10.34$ & $6.08$ & $13.36$ & $5.88$ & $%
5.58 $ \\ \hline
$-1.584\times 10^{-1}$ & $3.52$ & $10.76$ & $6.35$ & $13.41$ & $5.90$ & $%
5.62 $ \\ \hline
$-1.774\times 10^{-1}$ & $3.76$ & $11.00$ & $6.48$ & $13.40$ & $5.89$ & $%
5.60 $ \\ \hline\hline
&  &  &  &  &  &
\end{tabular}%
\end{center}
\end{table*}
\begin{table*}[tbp]
\caption{Structure properties of neutron star in massive gravity for $%
m^{2}c_{1}=3.168\times 10^{-5}$ and $m^{2}c_{2}=-3.168\times 10^{-2}$.}
\label{tab2}
\begin{center}
\begin{tabular}{ccccccc}
\hline\hline
$C$ & ${M_{max}}\ (M_{\odot})$ & $R\ (km)$ & $R_{Sch}\ (km)$ & $\overline{%
\rho }$ $(10^{14}g$ $cm^{-3})$ & $\sigma (10^{-1})$ & $z(10^{-1})$ \\
\hline\hline
$0.01$ & $1.68$ & $8.42$ & $4.20$ & $13.36$ & $5.00$ & $5.58$ \\ \hline
$0.10$ & $1.68$ & $8.42$ & $4.84$ & $13.36$ & $5.75$ & $5.58$ \\ \hline
$0.50$ & $1.70$ & $8.45$ & $4.96$ & $13.37$ & $5.87$ & $5.58$ \\ \hline
$1.00$ & $1.76$ & $8.55$ & $5.03$ & $13.37$ & $5.88$ & $5.58$ \\ \hline
$2.00$ & $2.01$ & $8.94$ & $5.26$ & $13.36$ & $5.88$ & $5.58$ \\ \hline
$3.00$ & $2.45$ & $9.55$ & $5.62$ & $13.36$ & $5.88$ & $5.58$ \\ \hline
$4.00$ & $3.11$ & $10.34$ & $6.08$ & $13.36$ & $5.88$ & $5.58$ \\ \hline
$4.73$ & $3.76$ & $11.00$ & $6.49$ & $13.40$ & $5.90$ & $5.61$ \\
\hline\hline
&  &  &  &  &  &
\end{tabular}%
\end{center}
\end{table*}
\begin{table*}[tbp]
\caption{Structure properties of neutron star in massive gravity for $C=2$
and $m^{2}c_{2}=-3.168\times 10^{-2}$.}
\label{tab3}
\begin{center}
\begin{tabular}{ccccccc}
\hline\hline
$m^{2}c_{1}$ & ${M_{max}}\ (M_{\odot})$ & $R\ (km)$ & $R_{Sch}\ (km)$ & $%
\overline{\rho }$ $(10^{14}g$ $cm^{-3})$ & $\sigma (10^{-1})$ & $z(10^{-1})$
\\ \hline\hline
$3.168\times 10^{-14}$ & $2.01$ & $8.94$ & $5.26$ & $13.36$ & $5.88$ & $5.58$
\\ \hline
$3.168\times 10^{-13}$ & $2.01$ & $8.94$ & $5.26$ & $13.36$ & $5.88$ & $5.58$
\\ \hline
$3.168\times 10^{-12}$ & $2.01$ & $8.94$ & $5.26$ & $13.36$ & $5.88$ & $5.58$
\\ \hline
$3.168\times 10^{-11}$ & $2.01$ & $8.94$ & $5.26$ & $13.36$ & $5.88$ & $5.58$
\\ \hline
$3.168\times 10^{-10}$ & $2.01$ & $8.94$ & $5.26$ & $13.36$ & $5.88$ & $5.58$
\\ \hline
$-3.168\times 10^{-14}$ & $2.01$ & $8.94$ & $5.36$ & $13.36$ & $5.99$ & $%
5.58 $ \\ \hline
$-1.584\times 10^{-13}$ & $2.01$ & $8.94$ & $5.27$ & $13.36$ & $5.89$ & $%
5.58 $ \\ \hline
$-3.168\times 10^{-13}$ & $2.01$ & $8.94$ & $5.26$ & $13.36$ & $5.88$ & $%
5.58 $ \\ \hline
$-3.168\times 10^{-12}$ & $2.01$ & $8.94$ & $5.26$ & $13.36$ & $5.88$ & $%
5.58 $ \\ \hline
$-3.168\times 10^{-11}$ & $2.01$ & $8.94$ & $5.26$ & $13.36$ & $5.88$ & $%
5.58 $ \\ \hline
$-3.168\times 10^{-10}$ & $2.01$ & $8.94$ & $5.26$ & $13.36$ & $5.88$ & $%
5.58 $ \\ \hline\hline
&  &  &  &  &  &
\end{tabular}%
\end{center}
\end{table*}

It is notable that, here, we ignore the effects of cosmological
constant on the structure of neutron star. For investigating its
effects, we refer the interested reader to Ref. \cite{TOV-Lambda},
in which, it was shown that, this constant has no effect on the
structure of this star when the cosmological constant is about
$10^{-52}\ m^{-2}$. Now, we are in a position to study the
properties of neutron star in massive gravity. First, we consider
the mass of graviton about $1.78\times 10^{-65}g$, which was
obtained in Ref. \cite{AliD}. Next, we use obtained results by A.
W. Steiner et al \cite{Steiner2010} in which an empirical dense
matter EoS from a heterogeneous data set of six neutron stars was
obtained. Their results showed that the radius of a neutron star
must be in the range of $R\leq \left( 11\thicksim 14\right) km$.
In the present paper, we consider the maximum radius of neutron
star in the range of $R\leq 11km$ and investigate the maximum mass
for neutron star in the massive gravity by employing the modern
EoS of neutron star matter derived from microscopic calculations.
According to the table \ref{tab1}, considering the spacial values
for the parameters of the modified TOV equation, the maximum mass
of neutron star is an increasing function of $m^{2}c_{2}$.
Calculations show that the maximum mass of neutron star can be
more than $3M_{\odot }$ ($M_{\max }\approx 3.8M_{\odot }$),
whereas in the Einstein gravity and by using this EoS, the maximum
mass was in the range of $M_{\max }\leq 1.68M_{\odot }$. A mass
measurement for PSR J1614-2230 \cite{PSRJ1} showed that the mass
for neutron star was about $2M_{\odot }$. In other words, our
results cover the mass measurement of massive neutron star, and
also, predict that the mass of neutron star in massive gravity can
be in the range upper than $3M_{\odot }$ (see the table \ref{tab1}
for more details). Also, by decreasing the value of $m^{2}c_{2}$
less than ($10^{-4}$), the maximum mass and radius of
neutron star are not affected. In other words, considering the value of $%
m^{2}c_{2}$ about $-10^{-4}$, the maximum mass and radius of neutron star
reduce to the obtained results of massless Einstein gravity \cite{BordbarH}.

On the other hand, the average density ($\overline{\rho }$) of the neutron
star calculated in the tables \ref{tab1}, \ref{tab2} and \ref{tab3} shows
that the central density may exceed a few times as $10^{15}\ g~cm^{-3}$. In
other words, it is larger than the normal nuclear density, $\rho
_{0}=2.7\times 10^{14}g~cm^{-3}$ \cite{Wiringa}.

For further investigation, we plot the mass of neutron star versus the
central mass density ($\rho _{c}$) in left panels of Figs. \ref{Fig3} and %
\ref{Fig4}. As one can see, the maximum mass of this star increases as $%
m^{2}c_{2}$ increases. On the other hand, the variation of maximum mass
versus radius is also shown in right panels of Figs. \ref{Fig3} and \ref%
{Fig4}. 

Now, we complete our discussion by considering the gravitational
mass equal to $1.78\times 10^{-65}g$, with various values for
different parameters of modified TOV equation (see Eq.
(\ref{4TOV})) and obtain the maximum mass of neutron star in the
massive gravity. The results are
presented in the tables \ref{tab2} and \ref{tab3}. According to the table %
\ref{tab2}, the maximum mass of neutron star is an increasing function of $C$%
. It is notable that considering the values less than $0.1$ for $C$, the
maximum mass and corresponding radius of neutron star are not affected. In
other words, these results reduce to the obtained results of the maximum
mass and radius of neutron star in the Einstein gravity \cite{BordbarH}. The
variation of $m^{2}c_{1}$ has very interesting effects. In this case, the
maximum mass and radius of this star are constant and by variation of $%
m^{2}c_{1}$, these quantities are not affected (see the table \ref{tab3}).

For completeness, in the following, we investigate other properties of
neutron star in this gravity such as the Schwarzschild radius, average
density, compactness, the gravitational redshift and dynamical stability.

\subsubsection{modified Schwarzschild Radius}

It is clear that by applying the massive term to the Einstein gravity, the
Schwarzschild radius is modified. Considering Eq. (\ref{4g(r)}) and using
the horizon radius constraint ($g(r)=0$), we can obtain the Schwarzschild
radius ($R_{Sch}$) for the EN-massive gravity. After some calculations, the
Schwarzschild radius for this gravity without the cosmological constant is
obtained as
\begin{equation}
R_{Sch}=\frac{c\left( 1-m^{2}c_{2}C^{2}\right) }{m^{2}cc_{1}C}-\frac{\sqrt{%
c^{2}\left( m^{2}c_{2}C^{2}-1\right) ^{2}-4m^{2}c_{1}CGM}}{m^{2}cc_{1}C}.
\label{Sch}
\end{equation}

Using the series expansion of $R_{Sch}$ for the limit $m^{2}\rightarrow 0$,
we find that
\begin{equation}
R_{Sch}\approx \frac{2GM}{c^{2}}+\frac{2GMC\left( c^{2}c_{2}C+c_{1}GM\right)
}{c^{4}}m^{2}+O(m^{4}),
\end{equation}%
where the first term is the Schwarzschild radius in Einstein gravity \cite%
{Schwarzschild}, as expected, and the second term indicates the massive
correction.

In order to investigate the effects of various parameters on the modified
Schwarzschild radius, one can look at the tables \ref{tab1}, \ref{tab2} and %
\ref{tab3}. As one can see in tables \ref{tab1} and \ref{tab2}, by
increasing the maximum mass and radius of neutron star, the
Schwarzschild radius increases and these stars are out of the
Schwarzschild radius. Also, considering the negative value of
$m^{2}c_{2}$ and increasing $m^{2}c_{2}$, the Schwarzschild radius
increases (see table \ref{tab1}). On the other
hand, by increasing $C$, the Schwarzschild radius increases (see table \ref%
{tab2}). In addition, considering the positive (negative) values of $%
m^{2}c_{1}$ and increasing (decreasing) $m^{2}c_{1}$, the Schwarzschild
radius almost does not change (see table \ref{tab3}).

\subsubsection{Average Density}

Now, using the maximum mass and radius obtained in the massive gravity, we
can calculate the average density of neutron star in $4-$dimensions as
\begin{equation}
\overline{\rho }=\frac{3M}{4\pi R^{3}},  \label{density}
\end{equation}%
where the results for variation of the massive parameters are presented in
the tables \ref{tab1}, \ref{tab2} and \ref{tab3}. Considering different
parameters introduced in this theory, the average density of this star is
almost the same. In other words, by variations of the different parameters,
the average density remains fixed.

\subsubsection{Compactness}

The compactness of a spherical object may be defined by the ratio of
Schwarzschild radius to radius of that object
\begin{equation}
\sigma =\frac{R_{Sch}}{R},
\end{equation}%
which may be indicated as the strength of gravity. For the massive gravity,
we obtain the values of $\sigma $ in the tables \ref{tab1}, \ref{tab2} and %
\ref{tab3}. For different values of $m^{2}c_{2}$ and $C$, the results show
that the strength of gravity is almost the same (see tables \ref{tab1} and %
\ref{tab2}). But, for different values of $m^{2}c_{1}$, there are two
interesting behaviors. A) considering the positive value of $m^{2}c_{1}$ and
increasing $m^{2}c_{1}$, the strength of gravity do not change. B)
considering the negative values of $m^{2}c_{1}$ and increasing $m^{2}c_{1}$,
the strength of gravity decreases and the strength of gravity is not
affected for $m^{2}c_{1}>-3.168\times 10^{-13}$ (see the table \ref{tab3}).

\subsubsection{Gravitational redshift}

Considering Eq. (\ref{4g(r)}) for vanishing $\Lambda $ and by using
definition of the gravitational redshift, we obtain this quantity in the
massive gravity as
\begin{equation}
z=\frac{1}{\sqrt{1-m^{2}C\left( \frac{c_{1}r}{2}+c_{2}C\right) -\frac{2GM}{%
c^{2}r}}}-1,
\end{equation}%
in which it reduces to the gravitational redshift in the Einstein gravity
when $m^{2}=0$. The results show that, the gravitational redshift of neutron
star is almost independent of different parameters. The gravitational
redshift of each compact object depends on its average density, so as one
can see, the average density for these stars are almost the same, therefore
the gravitational redshift of them must be the same.%
\begin{figure}[tbp]
$%
\begin{array}{cc}
\epsfxsize=10cm \epsffile{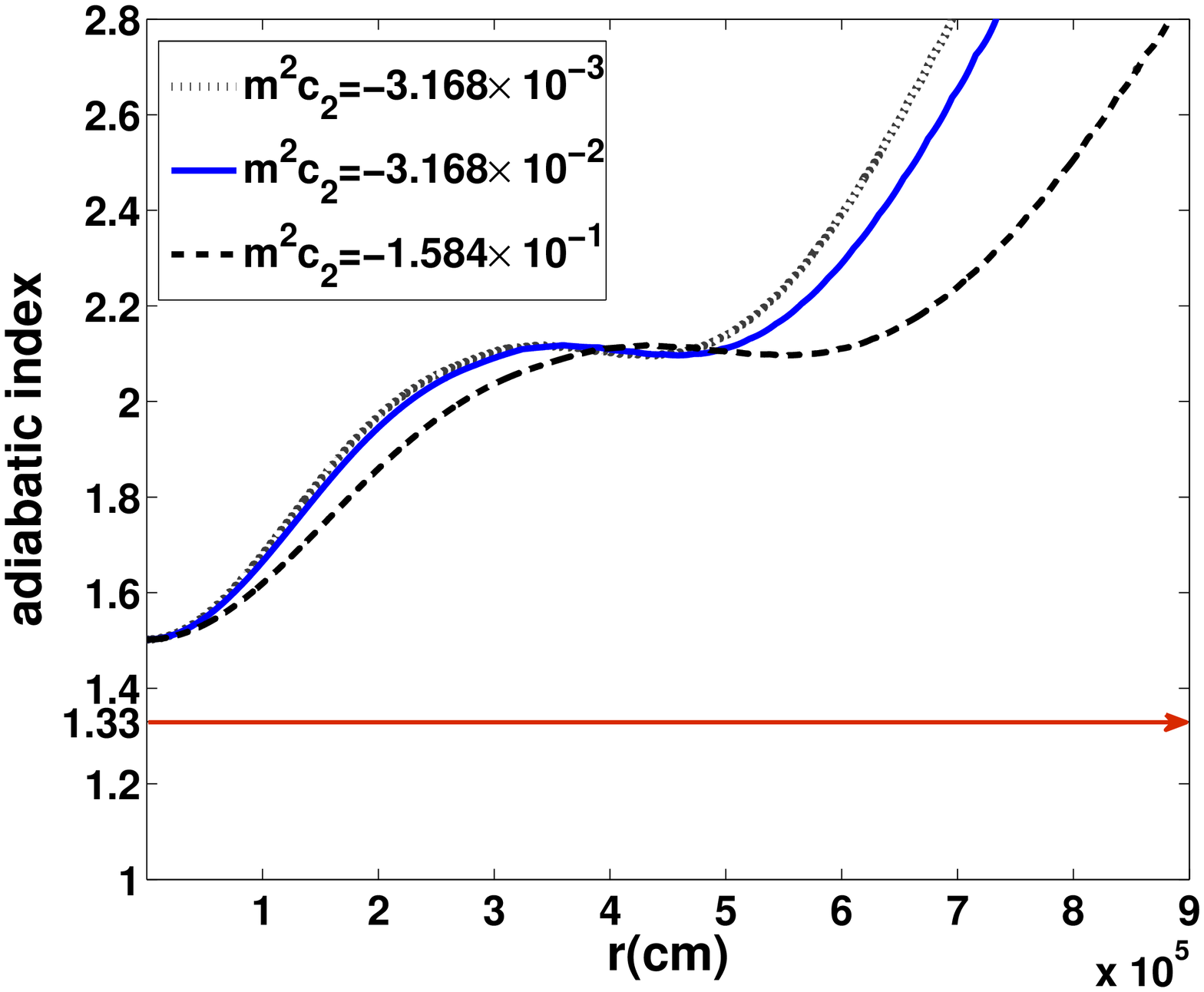} & \epsfxsize=10cm %
\epsffile{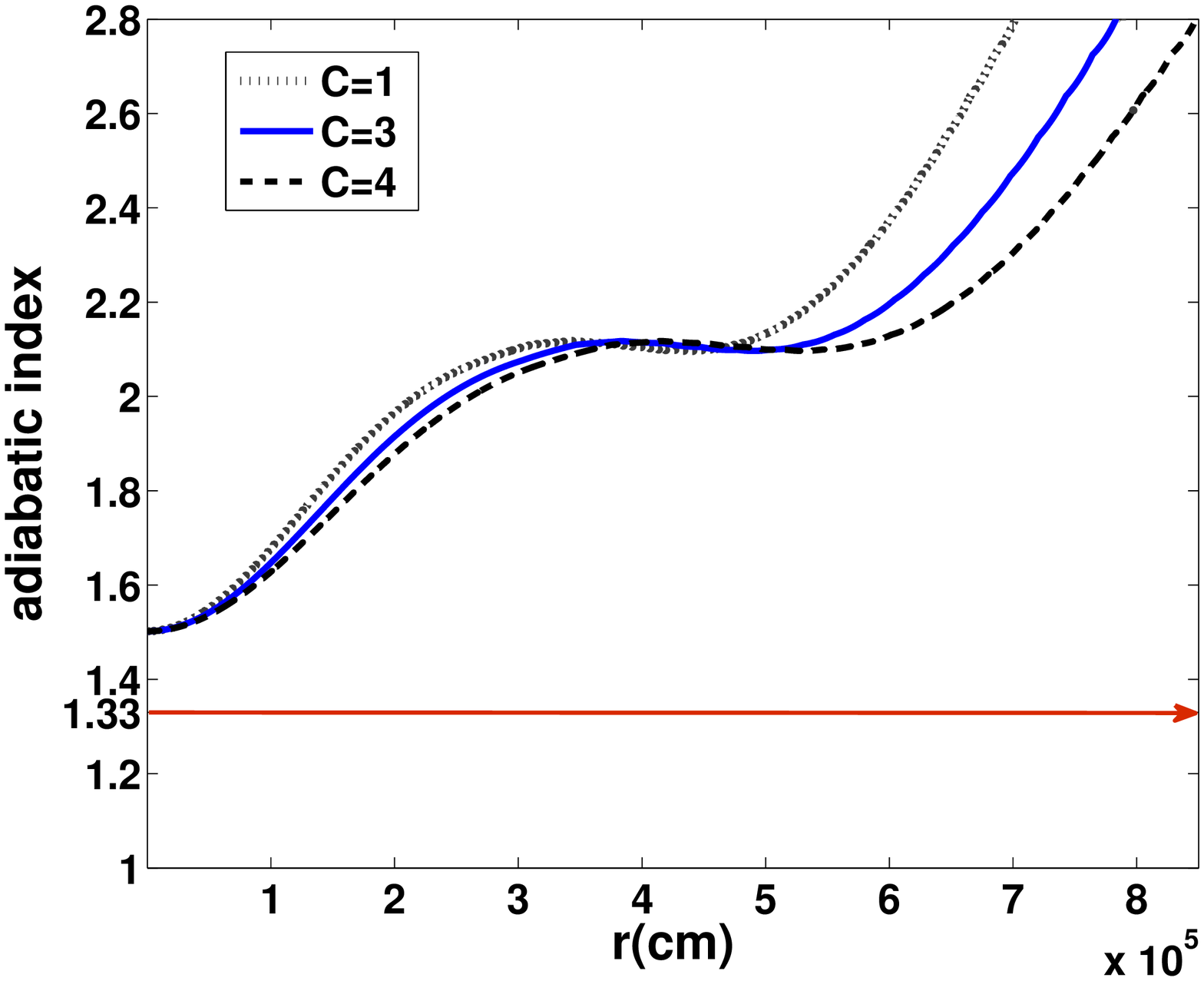}%
\end{array}
$%
\caption{Adiabatic index versus radius for $m^{2}c_{1}=3.168\times 10^{-13}$.%
\newline
Left diagrams: for $C=2$, $m^{2}c_{2}=-3.168\times 10^{-3}$ (doted line), $%
m^{2}c_{2}=-3.168\times 10^{-2}$ (continuous line) and $m^{2}c_{2}=-1.584%
\times 10^{-1}$ (dashed line). \newline
Right diagrams: for $m^{2}c_{2}=-3.168\times 10^{-2}$, $C=1.0$ (doted line),
$C=3.0$ (continuous line) and $C=4.0$ (dashed line).}
\label{Fig5}
\end{figure}


\begin{figure}[tbp]
$%
\begin{array}{cc}
\epsfxsize=10cm \epsffile{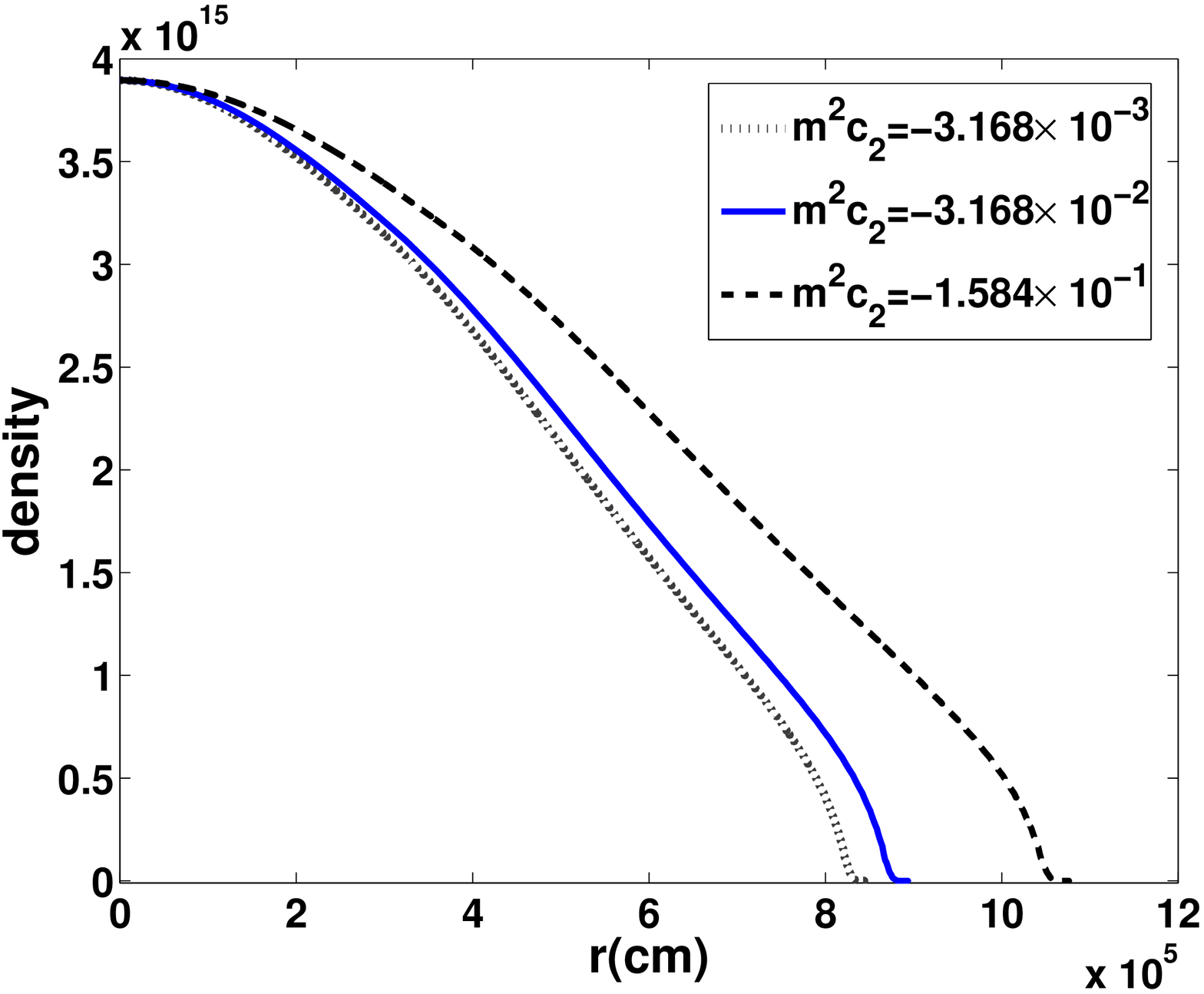} & \epsfxsize=10cm %
\epsffile{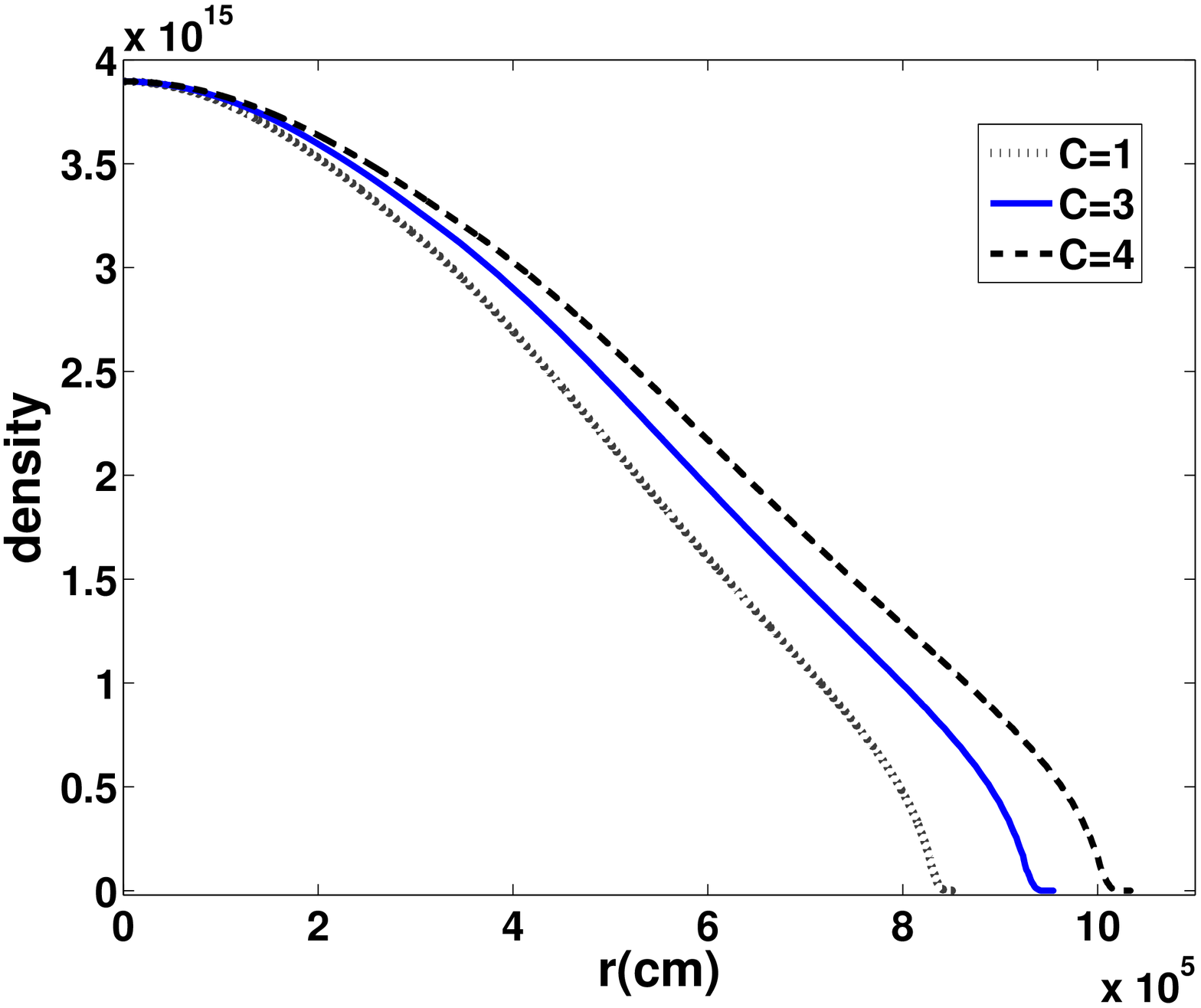}%
\end{array}
$%
\caption{Density versus radius for $m^{2}c_{1}=3.168\times 10^{-13}$.\newline
Left diagrams: for $C=2$, $m^{2}c_{2}=-3.168\times 10^{-3}$ (doted line), $%
m^{2}c_{2}=-3.168\times 10^{-2}$ (continuous line) and $m^{2}c_{2}=-1.584%
\times 10^{-1}$ (dashed line). \newline
Right diagrams: for $m^{2}c_{2}=-3.168\times 10^{-2}$, $C=1.0$ (doted line),
$C=3.0$ (continuous line) and $C=4.0$ (dashed line).}
\label{Fig6}
\end{figure}


\begin{figure}[tbp]
$%
\begin{array}{cc}
\epsfxsize=10cm \epsffile{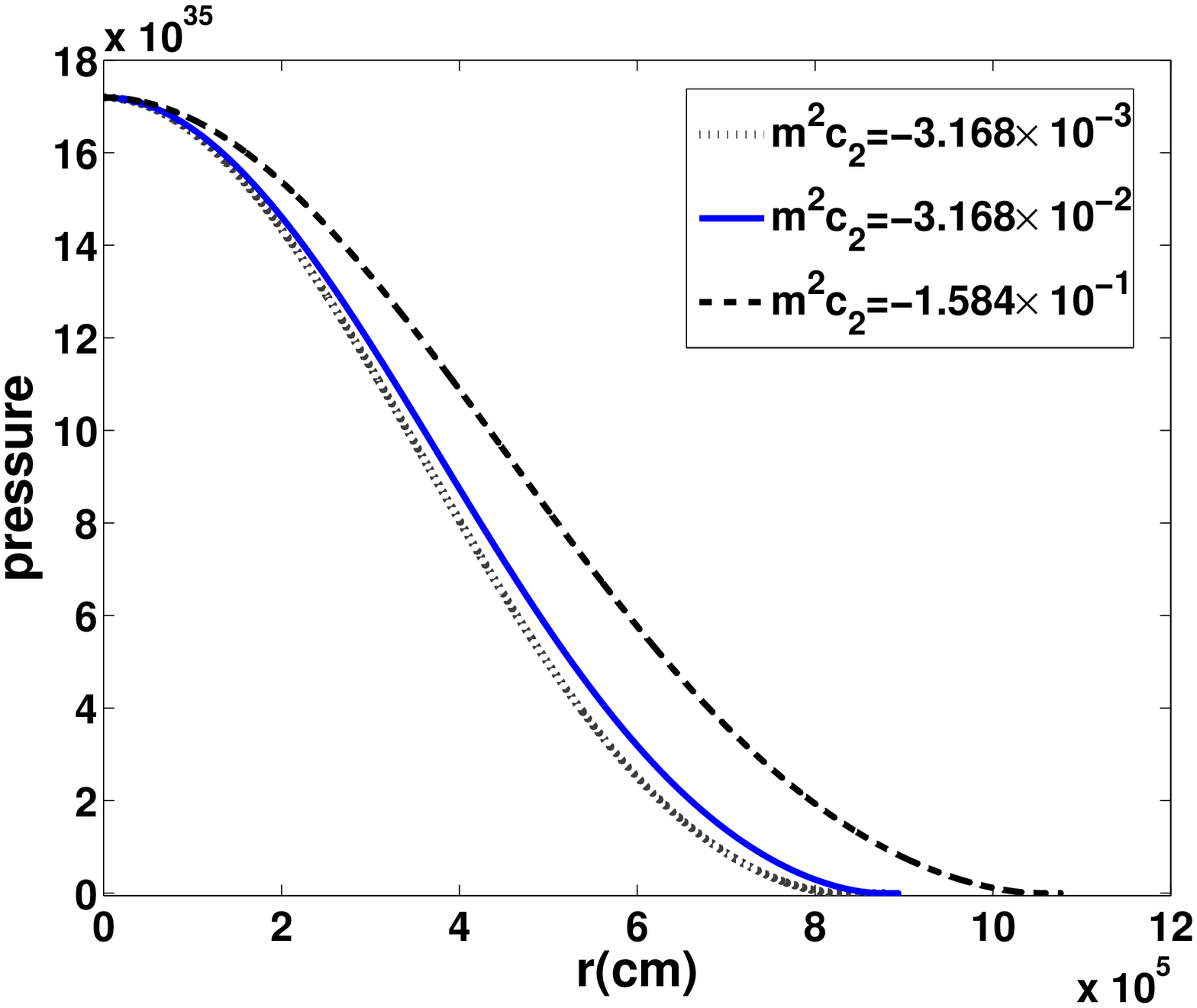} & \epsfxsize=10cm %
\epsffile{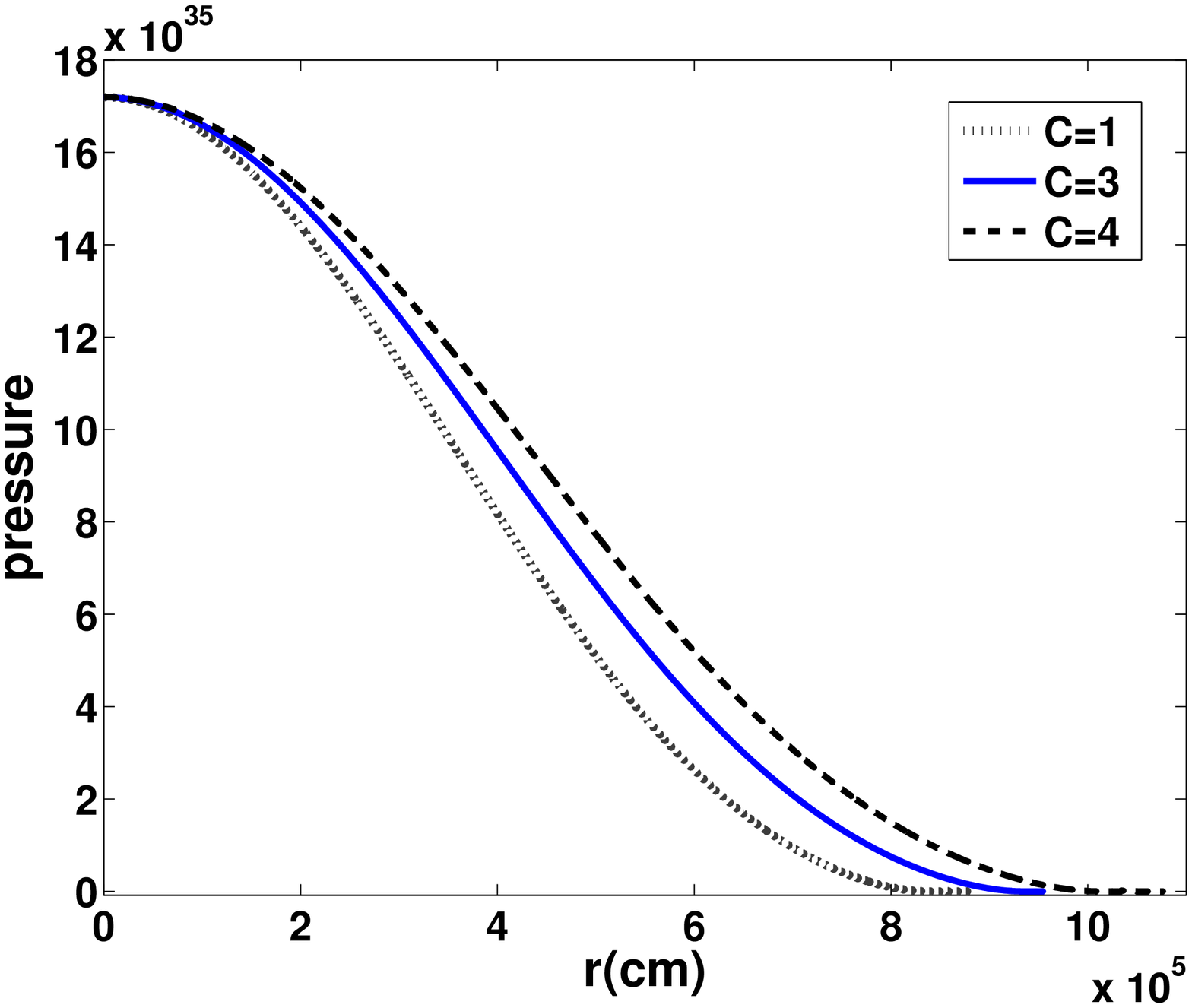}%
\end{array}
$%
\caption{Pressure versus radius for $m^{2}c_{1}=3.168\times 10^{-13}$.%
\newline
Left diagrams: for $C=2$, $m^{2}c_{2}=-3.168\times 10^{-3}$ (doted line), $%
m^{2}c_{2}=-3.168\times 10^{-2}$ (continuous line) and $m^{2}c_{2}=-1.584%
\times 10^{-1}$ (dashed line). \newline
Right diagrams: for $m^{2}c_{2}=-3.168\times 10^{-2}$, $C=1.0$ (doted line),
$C=3.0$ (continuous line) and $C=4.0$ (dashed line).}
\label{Fig7}
\end{figure}


\subsubsection{Dynamical Stability}

The dynamical stability of stellar model against infinitesimal radial
adiabatic perturbation was introduced by Chandrasekhar in Ref. \cite%
{Chandrasekhar}. This stability condition was developed and applied to
astrophysical cases by many authors \cite{BardeenTM,Kuntsem,Mak,Kalam}. The
adiabatic index ($\gamma $) is defined as
\begin{equation}
\gamma =\frac{\rho c^{2}+P}{c^{2}P}\frac{dP}{d\rho }.  \label{adiabatic}
\end{equation}

It is notable that, in order to have the dynamical stability, $\gamma $
should be more than $\frac{4}{3}$ ($\gamma >\frac{4}{3}=1.33$) everywhere
within the isotropic star. Therefore, we plot two diagrams related to $%
\gamma $ versus radius for different values of $m^{2}c_{2}$ and $C$ in Fig. %
\ref{Fig5}. As one can see, these stellar models in massive gravity are
stable against the radial adiabatic infinitesimal perturbations.

Also, we plot the density (pressure) versus distance from the center of
neutron star. As one can see, the density and pressure are maximum at the
center and they decrease monotonically towards the boundary (see Figs \ref%
{Fig6} and \ref{Fig7}).

\section{Neutron star properties via Planck mass \label{Planck}}

Here, our aim is to obtain the mass of neutron star according to the Planck
mass. The neutron stars are supported against the gravitational force by
degeneracy pressure of nucleons which is mainly related to the strong
repulsive inter-nucleons force. It is notable that the nucleon-nucleon
interaction is so strong, and it is taken place through the pion exchange.
Therefore, we can consider the average density of a neutron star in term of
the nucleus density using the following form (see Refs. \cite%
{HendiBEP,Burrows} for more details)
\begin{equation}
\rho _{nuc}\sim \frac{3m_{p}}{4\pi \lambda _{\pi }^{3}},
\end{equation}%
where $m_{p}$ and $\lambda _{\pi }=\frac{\hbar }{m_{\pi }c}$ are,
respectively, the proton mass and Compton wavelength ($m_{\pi }$ is the pion
mass). Now, we are going to use an analogy for obtaining a relation between
the mass of neutron star in massive gravity and the Planck mass. Using the
equations (\ref{Sch}) and (\ref{density}) and by considering $\rho _{nuc}$,
one can derive the following corresponding mass as%
\begin{eqnarray}
\frac{3m_{p}}{4\pi \lambda _{\pi }^{3}} &\sim &\frac{3M}{4\pi R_{Sch}^{3}},
\nonumber \\
&&  \nonumber \\
M &\sim &\frac{1}{128m^{2}c_{1}Cm_{p}^{2}m_{\pi }^{6}G^{3}}\left\{ \left[
\mathcal{A}_{1}+8cm_{p}m_{\pi }^{3}G\right] \right. \sqrt{%
32m^{4}c_{1}^{2}C^{2}\hbar ^{3}\left( \frac{m^{4}c_{1}^{2}C^{2}\hbar ^{3}}{32%
}-\mathcal{A}_{2}\right) }  \nonumber \\
&&\left. +24m^{4}c_{1}^{2}C^{2}\hbar ^{3}\left[ m^{2}C^{2}\mathcal{A}%
_{3}-cm_{p}m_{\pi }^{3}G\right] \right\} ,  \label{PlackI}
\end{eqnarray}%
where%
\begin{eqnarray*}
\mathcal{A}_{1} &=&m^{2}C^{2}\left( m^{2}c_{1}^{2}\hbar
^{3}-8cc_{2}m_{p}m_{\pi }^{3}G\right) , \\
\mathcal{A}_{2} &=&cGm_{p}m_{\pi }^{3}\left( m^{2}c_{2}C^{2}-1\right) , \\
\mathcal{A}_{3} &=&cc_{2}m_{p}m_{\pi }^{3}G-\frac{m^{2}c_{1}^{2}\hbar ^{3}}{%
24}.
\end{eqnarray*}

Now, we use the relation between the proton (pion) mass and the Planck mass
\cite{Burrows} to obtain the mass of neutron star with respect to the Planck
mass%
\begin{equation}
m_{p}=\frac{m_{pl}}{\eta _{p}}\ \ \ \ \&\ \ \ \ \ m_{\pi }=\frac{m_{pl}}{%
\eta _{\pi }},  \nonumber
\end{equation}%
\begin{eqnarray}
M &\sim &\frac{3\eta _{p}\eta _{\pi }^{3}}{16m^{2}m_{pl}^{8}c_{1}CG^{3}}%
\left\{ \left[ \frac{m^{4}c_{1}^{2}C^{2}\hbar ^{3}\eta _{p}\eta _{\pi }^{3}}{%
24}-\mathcal{B}_{1}\right] \right. \sqrt{\frac{32m^{4}c_{1}^{2}C^{2}\hbar
^{3}}{\eta _{p}\eta _{\pi }^{3}}\left( \frac{m^{4}c_{1}^{2}C^{2}\hbar
^{3}\eta _{p}\eta _{\pi }^{3}}{32}-3\mathcal{B}_{1}\right) }  \nonumber \\
&&  \nonumber \\
&&\left. +m^{4}c_{1}^{2}C^{2}\hbar ^{3}\left[ \frac{-m^{4}c_{1}^{2}C^{2}%
\hbar ^{3}\eta _{p}\eta _{\pi }^{3}}{24}+3\mathcal{B}_{1}\right] \right\} ,
\label{PlanckII}
\end{eqnarray}%
where%
\begin{equation*}
\mathcal{B}_{1}=\frac{cm_{pl}^{4}G\left( m^{2}c_{2}C^{2}-1\right) }{3}.
\end{equation*}

It is notable that in the absence of massive term ($m^{2}\rightarrow 0$),
the obtained relation reduces to the usual general relativity case
\begin{equation}
M \sim \left( \frac{\hbar c}{G}\right) ^{3/2}\frac{1}{m_{p}^{2}}\left( \frac{%
\eta _{\pi }}{2\eta _{p}}\right) ^{3/2} \sim m_{pl}\eta _{p}^{2}\left( \frac{%
\eta _{\pi }}{2\eta _{p}}\right) ^{3/2}.
\end{equation}

Now, we are in a position to obtain a constraint on the neutron star radius.
Regarding the fact that the radius of neutron star should be greater than
the Schwarzschild radius, one can extract a limitation for the radius of
neutron star via the Planck mass as a fundamental physical constant. Using
Eq. (\ref{PlanckII}) with $R_{NS}>R_{Sch}$, we obtain
\begin{eqnarray}
R_{NS} &>&\frac{1}{8m^{2}cc_{1}m_{pl}^{4}C}\left\{ 8\sqrt{2}%
cGm_{pl}^{4}\left( m^{2}c_{2}C^{2}-1\right) \left( -m^{2}c_{1}C\hbar \eta
_{\pi }\left[ m^{4}c_{1}^{2}C^{2}\hbar ^{3}\eta _{p}\eta _{\pi }^{3}-24%
\mathcal{B}_{1}\right] \right. \right.  \nonumber \\
&&\times \sqrt{\hbar \eta _{p}\eta _{\pi }\left[ m^{4}c_{1}^{2}C^{2}\hbar
^{3}\eta _{p}\eta _{\pi }^{3}-96\mathcal{B}_{1}\right] }+\left(
m^{4}c_{1}^{2}C^{2}\hbar ^{3}\eta _{p}\eta _{\pi }^{3}\right)
^{2}-72m^{4}c_{1}^{2}C^{2}\hbar ^{3}\eta _{p}\eta _{\pi }^{3}\mathcal{B}_{1}
\nonumber \\
&&\left. \left. +32c^{2}G^{2}m_{pl}^{8}\left( m^{2}c_{2}C^{2}-1\right)
^{2}\right) ^{1/2}\right\} .  \label{Rlimit}
\end{eqnarray}

As a final comment, we should note that in the absence of the massive term,
Eq. (\ref{Rlimit}) reduces to
\begin{equation}
R_{NS}>G\left( \frac{\eta _{p}}{2\eta _{\pi }}\right) ^{1/2}\left( \frac{%
\eta _{\pi }}{cm_{pl}}\right) ^{2}\left( \frac{\hbar c}{G}\right) ^{3/2},
\end{equation}%
which may indicate a minimum value for neutron star radius in usual general
relativity.

\section{Closing Remarks \label{CONCLUSION}}

In this paper, we considered the spherically symmetric metric and extracted
a modified TOV equation of stars in the Einstein-massive gravity in $4-$%
dimensions. Then, we showed that for $m\rightarrow 0$ limit, the obtained
TOV in Einstein-massive gravity reduces to the Einstein-$\Lambda $ gravity.
The generalization of modified TOV equation to arbitrary $d-$dimensions was
also done (see appendix A). Furthermore, we have considered an EoS, which
was derived from microscopic calculations and investigated Le Chatelier's
principle for the mentioned EoS. It was shown that this equation is suitable
for investigating the structure of neutron star.

Considering the modified TOV obtained in this paper, the structure
of neutron star was investigated. The results showed that, the
maximum mass of these stars increases when $m^{2}c_{2}$ and $C$
increase (the results represented in various tables numerically).
In addition, it was shown that by considering the constant values
of $C$ and $m^{2}c_{2}$, the maximum mass of neutron star is
independent of $m^{2}c_{1}$.

Then, we showed that by increasing the maximum mass of neutron star, the
radius and the Schwarzschild radius increase as well. It is notable that, by
regarding massive graviton, the Schwarzschild radius is modified. In order
to conduct more investigations, we plotted some diagrams related to the
mass-radius and mass-central mass density. We found that these figures are
similar to the diagrams related to the mass-radius and mass-central mass
density in usual GR. In addition, these diagrams confirmed the validity of
obtained results in massive gravity.

After that the adiabatic index was investigated. It was shown that
this star is dynamically stable. It is notable that the density
and pressure are maximum at the center of the star and decrease
monotonically towards the boundary.

Jacoby et al \cite{Jacoby}\ and Verbiest et al \cite{Verbiest} used the
detection of Shapiro delay to measure the masses of both the neutron star
and its binary component. Also, using the same approach, the masses of
compact objects were obtained for Vela X-1 (about $1.8M_{\odot }$) \cite%
{Vela}, PSR J1614-2230 (about $1.97M_{\odot }$) \cite{PSRJ1}, PSR J0348+0432
(about $2.01M_{\odot }$) \cite{J0348}, 4U 1700-377 (about $2.4M_{\odot }$)
\cite{4U1700} and J1748-2021B (about $2.7M_{\odot }$) \cite{J1748}.\ It is
notable that, in this paper, we showed that the obtained maximum mass of
neutron star in massive gravity can cover all the measured masses of pulsars
and neutron stars. Also, we predicted the existence of possible mass of more
than $3M_{\odot}$.

Briefly, we obtained the quite interesting results from massive gravity for
the neutron star such as:

I) Obtaining the modified TOV equation. II) Prediction of maximum mass for
neutron star more than $3M_{\odot }$ $\left( M_{\max }\approx 3.8M_{\odot
}\right) $, due to the existence of massive gravitons. III) Dynamically
stable neutron star in the massive gravity. IV) EoS derived from microscopic
calculations satisfied the energy, stability conditions and Le Chatelier's
principle, simultaneously. V) The Schwarzschild radius was modified in the
presence of massive gravity. VI) Due to the considering massive graviton,
the gravitational redshift was modified. VII) The relations between the mass
and the radius of neutron star versus the Planck mass as a fundamental
physical constant were extracted. VIII) Our consequences covered previous
results and reduce to the Einstein gravity for massless graviton ($m=0$), as
expected.

Finally, it is notable that the investigation of other compact objects such
as quark star and white dwarf in the context of massive gravity and its
modified TOV equation are interesting subjects. Moreover, it is worth
studying the effects of higher dimensions and other equation of states on
the structure of compact objects. Also, anisotropic compact objects \cite%
{HarkoM,BoehmerH,SunzuM,PaulD,NgubelangaM,MauryaG,RatanpalTP}, rotating,
slowly rotating \cite%
{slowly1,slowly2,slowly3,slowly4,slowly5,slowly6,slowly7}, rapidly
rotating \cite{rapid1,rapid2,rapid3,rapid4,rapid5,rapid6} neutron
stars and obtain the Buchdahl limit
\cite{BI,Buchdahl1,Buchdahl2,Buchdahl3,BII,BIII,Buchdahl4} in the
context of massive gravity are interesting topics. Furthermore,
regarding the considerable effects of free parameters on the
existence of tachyon-like instabilities, it will be useful to
address the mentioned substantial instability. We leave these
issues for the future works.


\section{Acknowledgements}

The author wish to thank Shiraz University Research Council. This work has
been supported financially by the Research Institute for Astronomy and
Astrophysics of Maragha, Iran.

\begin{center}
\textbf{Appendix: Modified TOV equation in higher dimension }
\end{center}

Here, we are interested in obtaining the modified TOV equation in
Einstein-massive gravity in higher dimensions. So, we consider a spherical
symmetric space-time in higher dimensions as
\begin{equation}
ds^{2}=f(r)dt^{2}-g^{-1}(r)dr^{2}-r^{2}h_{ij}dx_{i}dx_{j},
\label{Metrichigher}
\end{equation}%
where $i,j=1,2,3,...,d-2$, and also $h_{ij}dx_{i}dx_{j}$ is the line element
of a $(d-2)-$dimensional unit sphere
\begin{equation}
h_{ij}dx_{i}dx_{j}=d\theta
_{1}^{2}+\sum\limits_{i=2}^{d-2}\prod\limits_{j=1}^{i-1}\sin ^{2}\theta
_{j}d\theta _{i}^{2}.
\end{equation}

We also use the following ansatz for the reference metric which is
introduced in Ref. \cite{CaiHPZ}
\begin{equation}
f_{\mu \nu }=diag(0,0,C^{2}r^{2}h_{ij}).  \label{fd1}
\end{equation}

Using the mentioned information and ansatz, we can find the explicit
functional forms of $\mathcal{U}_{i}$'s as
\begin{eqnarray}
\mathcal{U}_{1} &=&\frac{d_{2}C}{r},\text{ \ \ }\mathcal{U}_{2}=\frac{%
d_{2}d_{3}C^{2}}{r^{2}},  \nonumber \\
\mathcal{U}_{3} &=&\frac{d_{2}d_{3}d_{4}C^{3}}{r^{3}},\text{ \ }\mathcal{U}%
_{4}=\frac{d_{2}d_{3}d_{4}d_{5}C^{4}}{r^{4}},
\end{eqnarray}%
where we denoted $d_{i}=d-i$. We can also obtain the nonzero components of
the energy-momentum for $d$-dimensional perfect fluid as
\begin{equation}
T_{0}^{0}=\rho c^{2}~\ \&\ \
T_{1}^{1}=T_{2}^{2}=T_{3}^{3}=...=T_{d-1}^{d-1}=-P.  \label{higherdim}
\end{equation}

Considering the metric (\ref{Metrichigher}) with Eq. (\ref{higherdim}), we
can find the components of Eq. (\ref{Field equation}) are calculated as
\begin{eqnarray}
K_{d}c^{2}r^{2}\rho &=&\Lambda r^{2}+\frac{d_{2}d_{3}}{2}\left( 1-g\right) -%
\frac{d_{2}}{2}r g{^{\prime
}-}\frac{m^{2}d_{2}C}{2r^{d_{4}}}\left[
c_{1}r^{d_{3}}+d_{3}c_{2}Cr^{d_{4}}\right.  \nonumber \\
&&\left. +d_{3}d_{4}\left(
c_{3}C^{2}r^{d_{5}}+d_{5}c_{4}C^{3}r^{d_{6}}\right) \right] ,  \label{111} \\
&&  \nonumber \\
K_{d}fr^{2}P &=&-\Lambda r^{2}f-\left( 1-g\right) f+rg{f{^{\prime }}}+\frac{%
m^{2}d_{2}C}{2r^{d_{4}}}\left[ c_{1}r^{d_{3}}+d_{3}c_{2}Cr^{d_{4}}\right.
\nonumber \\
&&\left. +d_{3}d_{4}\left(
c_{3}C^{2}r^{d_{5}}+d_{5}c_{4}C^{3}r^{d_{6}}\right) \right] ,  \label{222} \\
&&  \nonumber \\
4K_{d}f^{2}r P &=&-4\Lambda rf^{2}+2\left( gf\right) {^{\prime }}f-r g f{%
^{\prime 2}}+r\left[ g{^{\prime }}f{^{\prime }+2g}f{^{\prime \prime }}\right]
f+\frac{2m^{2}d_{3}Cf^{2}}{r^{d_{4}}}\left[
c_{1}r^{d_{4}}+d_{4}c_{2}Cr^{d_{5}}\right.  \nonumber \\
&&\left. +d_{4}d_{5}\left( c_{3}C^{2}r^{d_{6}}+c_{4}C^{3}r^{d_{7}}\right)
\right] .  \label{333}
\end{eqnarray}

Considering Eqs. (\ref{111})-(\ref{333}) and after some calculations, one
can find a relation which is the same as Eq. (\ref{extraEQ}). In addition,
we can obtain the functional form of $g(r)$ by using Eq. (\ref{111}) as
\begin{equation}
g(r)=1+\frac{2\Lambda }{d_{1}d_{2}}r^{2}-\frac{c^{2}K_{d}M(r)\Gamma \left(
\frac{d_{1}}{2}\right) }{d_{2}\pi ^{d_{1}/2}r^{d_{3}}}-m^{2}\left[ \frac{%
c_{1}C}{d_{2}}r+c_{2}C^{2}+\frac{d_{3}c_{3}C^{3}}{r}+\frac{%
d_{3}d_{4}c_{4}C^{4}}{r^{2}}\right] ,  \label{dg(r)}
\end{equation}%
where $M(r)=\int \frac{2\pi ^{d_{1}/2}}{\Gamma (d_{1}/2)}r^{d_{2}}\rho (r)dr$
and $\Gamma $ is the gamma function, which satisfies some conditions such as
$\Gamma (1/2)=\sqrt{\pi }$, $\Gamma (1)=1$ and $\Gamma (x+1)=x\Gamma (x)$.

Now, we obtain $f^{\prime }$ from Eq. (\ref{222}) and insert it with Eq. (%
\ref{dg(r)}) into Eq. (\ref{extraEQ}) to obtain the following higher
dimensional HEE in Einstein-massive gravity
\begin{eqnarray}
\frac{dP}{dr} &=&\left( c^{2}\rho +P\right) \left\{ \frac{\left[ \frac{d_{3}%
}{2}\left[ g(r)-1\right] -\frac{r^{2}}{d_{2}}\left( \Lambda +K_{d}P\right) +%
\right] }{r g}\right. +\frac{m^{2}C}{2gr^{d_{3}}}\left[ \left(
c_{1}r^{d_{3}}+d_{3}c_{2}Cr^{d_{4}}\right. \right.  \nonumber \\
&&\left. \left. \left. +d_{3}d_{4}C^{2}\left(
c_{3}r^{d_{5}}+d_{5}c_{4}Cr^{d_{6}}\right) \right) \right] \right\} ,
\label{TOVd}
\end{eqnarray}%
where $g(r)$ is presented in Eq. (\ref{dg(r)}).

As a special case, it is notable that for $m=0$, Eq. (\ref{TOVd}) reduces to
the following $d$-dimensional TOV equation obtained in Einstein-$\Lambda $
gravity \cite{TOV-Lambda}%
\begin{equation}
\frac{dP}{dr}=\frac{\left( c^{2}\rho +P\right) \left[ \frac{(d-1)(d-3)\Gamma
(\frac{d-1}{2})}{4\pi ^{\left( d-1\right) /2}}c^{2}K_{d}M(r)+r^{d-1}\left(
\Lambda +\frac{d-1}{2}K_{d}P\right) \right] }{r\left[ -\Lambda
r^{d-1}+(d-1)\left( \frac{\Gamma (\frac{d-1}{2})}{2\pi ^{^{\left( d-1\right)
/2}}}c^{2}K_{d}M(r)-\frac{d-2}{2}r^{d-3}\right) \right] }.  \label{TOVLambda}
\end{equation}

Regarding a suitable EoS for higher dimensional spacetime, with
the obtained modified $d$-dimensional TOV equation, one can
investigate the neutron stars in higher dimensional massive
gravity. We leave the mentioned problem for the future works.

\begin{center}
\textbf{Appendix: B brief dimensional analysis of massive parameters and its
values}
\end{center}

Here, we are going to investigate the massive coefficients via dimensional
analysis. In general, all terms of Eq. (\ref{dg(r)}), including $m^{2}\frac{%
c_{1}C}{d_{2}}r$, $m^{2}c_{2}C^{2}$, $\frac{m^{2}d_{3}c_{3}C^{3}}{r}$ and $%
\frac{m^{2}d_{3}d_{4}c_{4}C^{4}}{r^{2}}$ must be dimensionless. On the other
hand, in dimensional analysis we know that $[m]=M$, $[r]=L$ and $[d_{n}]=1$.
Therefore, the dimensional interpretation of massive terms are%
\begin{eqnarray}
\lbrack c_{1}C] &=&M^{-2}L^{-1},  \label{c1} \\
\lbrack c_{2}C^{2}] &=&M^{-2},  \label{c2} \\
\lbrack c_{3}C^{3}] &=&M^{-2}L,  \label{c3} \\
\lbrack c_{4}C^{4}] &=&M^{-2}L^{2}.  \label{c4}
\end{eqnarray}

Using Eqs. (\ref{c1})-(\ref{c4}), one can show that massive coefficients
are, dimensionaly,%
\begin{equation*}
\lbrack C]=L\text{ \ \ \ \ \ \ \&\ \ \ \ \ \ \ \ }[c_{i}]=M^{-2}L^{-2},\ \ \
\ i=1,2,3,4
\end{equation*}

On the other hand, regarding the dimensionless action (\ref{Action}), we
find that dimensional interpretations of all $R$, $2\Lambda $ and $%
m^{2}\sum_{i}^{4}c_{i}U_{i}(g,f)$ are $L^{-2}$. Remembering that $%
[m^{2}c_{i}]=L^{-2}$, one can conclude that $U_{i}$'s are dimensionless. In
addition, like the cosmological constant, $m^{2}c_{i}$ terms could play the
role of the pressure in the extended phase space (see Ref. \cite{Kubiznak}
to find details regarding the relation between pressure and the cosmological
constant).




\end{document}